\documentclass[aps,prx,amsmath,amssymb,notitlepage,superscriptaddress,longbibliography, twocolumn]{revtex4-1}
\usepackage[utf8]{inputenc}
\usepackage{amsmath}
\usepackage{graphicx}
\usepackage{dcolumn}
\usepackage{bm}
\usepackage{makeidx}
\usepackage[alsoload=synchem]{siunitx} 
\usepackage{physics}
\usepackage[usenames, dvipsnames]{xcolor}
\usepackage[colorlinks=true, citecolor={blue!60!black}, urlcolor={blue!60!black}, linkcolor = {blue!50!black}]{hyperref}
\usepackage{cleveref}	
\usepackage{enumitem}
\usepackage{gensymb}
\usepackage[nolist,nohyperlinks]{acronym}
\usepackage{fancyhdr}
\usepackage{svg}
\usepackage{siunitx}

\makeatletter
\def\l@subsubsection#1#2{}
\makeatother

\begin{document}
\title{Protocol to identify a topological superconducting phase in a three-terminal device}

\author{Dmitry~I.~Pikulin}
\affiliation{Microsoft Quantum, Station Q, University of California, Santa Barbara, CA 93106, USA}
\affiliation{Microsoft Quantum, Redmond, Washington 98052, USA}

\author{Bernard~van~Heck}
\affiliation{Microsoft Quantum, Station Q, University of California, Santa Barbara, CA 93106, USA}
\affiliation{Microsoft Quantum Lab Delft, Delft University of Technology, 2600 GA Delft, The Netherlands}

\author{Torsten~Karzig}
\affiliation{Microsoft Quantum, Station Q, University of California, Santa Barbara, CA 93106, USA}

\author{Esteban~A.~Martinez}
\affiliation{Center for Quantum Devices, Niels Bohr Institute,
	University of Copenhagen, and Microsoft Quantum - Copenhagen,
	Universitetsparken 5, 2100 Copenhagen, Denmark}

\author{Bas~Nijholt}
\affiliation{Microsoft Quantum Lab Delft, Delft University of Technology, 2600 GA Delft, The Netherlands}

\author{Tom~Laeven}
\affiliation{Microsoft Quantum Lab Delft, Delft University of Technology, 2600 GA Delft, The Netherlands}

\author{Georg~W.~Winkler}
\affiliation{Microsoft Quantum, Station Q, University of California, Santa Barbara, CA 93106, USA}

\author{John~D.~Watson}
\affiliation{Microsoft Quantum Lab Delft, Delft University of Technology, 2600 GA Delft, The Netherlands}

\author{Sebastian~Heedt}
\affiliation{Microsoft Quantum Lab Delft, Delft University of Technology, 2600 GA Delft, The Netherlands}

\author{Mine~Temurhan}
\affiliation{Microsoft Quantum Lab Delft, Delft University of Technology, 2600 GA Delft, The Netherlands}

\author{Vicky~Svidenko}
\affiliation{Microsoft Quantum, Redmond, Washington 98052, USA}

\author{Roman~M.~Lutchyn}
\affiliation{Microsoft Quantum, Station Q, University of California, Santa Barbara, CA 93106, USA}

\author{Mason~Thomas}
\affiliation{Microsoft Quantum, Station Q, University of California, Santa Barbara, CA 93106, USA}

\author{Gijs~de~Lange}
\affiliation{Microsoft Quantum Lab Delft, Delft University of Technology, 2600 GA Delft, The Netherlands}

\author{Lucas~Casparis}
\affiliation{Center for Quantum Devices, Niels Bohr Institute,
	University of Copenhagen, and Microsoft Quantum - Copenhagen,
	Universitetsparken 5, 2100 Copenhagen, Denmark}

\author{Chetan~Nayak}
\affiliation{Microsoft Quantum, Station Q, University of California, Santa Barbara, CA 93106, USA}
\affiliation{Department of Physics, University of California, Santa Barbara, CA 93106, USA}

\date{\today}

\begin{abstract}
	We develop a protocol to determine the presence and extent of a topological phase with Majorana zero modes in a hybrid superconductor-semiconductor device.
	The protocol is based on conductance measurements in a three-terminal device with two normal leads and one superconducting lead.
	A radio-frequency technique acts as a proxy for the measurement of local conductance, allowing a rapid, systematic scan of the large experimental phase space of the device.
	Majorana zero modes cause zero bias conductance peaks at each end of the wire, so we identify promising regions of the phase space by filtering for this condition.
	To validate the presence of a topological phase, a subsequent measurement of the non-local conductance in these regions is used to detect a topological transition via the closing and reopening of the bulk energy gap.
	We define data analysis routines that allow for an automated and unbiased execution of the protocol.
	Our protocol is designed to screen out false positives, especially trivial Andreev bound states that mimic Majorana zero modes in local conductance.
	We apply the protocol to several examples of simulated data illustrating the detection of topological phases and the screening of false positives.
\end{abstract}

\maketitle

\section{Introduction}

Topological quantum computing proposes to encode qubits in exponentially protected quantum states~\cite{freedman2003topological}.
The protected degrees of freedom are non-Abelian anyons with non-trivial braiding statistics~\cite{nayak2008non,TQCreview2015}.
Exchanging the positions of the anyons or, equivalently, performing a carefully chosen sequence of measurements~\cite{bonderson2008measurement} realizes quantum operations on the qubits.
Majorana zero-energy modes (MZMs) bound to vortices or other defects are one of the simplest types of non-Abelian anyons~\cite{read2000paired,ivanov2001non}.
MZMs correspond to neutral excitations of a topological superconductor consisting of an equal superposition of an electron and a hole.
Two well-separated MZMs form a single non-local fermionic state~\cite{kitaev2001unpaired}, allowing the encoding of quantum information which is immune to local bosonic perturbations (up to exponentially small corrections) and, thus, enabling the realization of qubits with long coherence times and high-fidelity operations.
A number of practical proposals for realizing topological superconductivity and MZMs have been put forward recently~\cite{fu2008superconducting, fu2009josephson, sau2010generic, lutchyn2010majorana, oreg2010helical,Hasan_2010, Alicea_2012,lutchyn2018majorana, flensberg2021engineered}.
A promising scalable approach to topological quantum computing is possible by engineering a quasi-1D network of topological superconductors supporting Majorana zero modes (MZMs)
~\cite{Alicea_2011,Fidkowski_2011,Clarke_2011,hyart2013flux,mi2013proposal,Aasen_2016,plugge2017majorana,Vijay_2016,karzig2017scalable,manousakis2017majorana,pikulin2020proposal}.
To harness this potential, we need to reliably identify a topological superconducting state and manipulate well-separated MZMs.
The purpose of this work is thus to devise practical criteria to identify MZMs in experimental systems with high certainty and formulate a protocol to test whether a device fulfills these criteria.

Several experimental systems were predicted to host MZMs.
A non-exhaustive list includes vortices in $p+ip$ superfluids~\cite{kopnin1991mutual,volovik2003universe,read2000paired,ivanov2001non} as well as topological insulators~\cite{fu2008superconducting, fu2009josephson}, semiconductors~\cite{sau2010generic, lutchyn2010majorana, oreg2010helical, stanescu2011majorana,Hell_2017, Pientka_2017, vaitiekenas2020fullshell} and magnetic atomic chains~\cite{choy2011majorana,Klinovaja_2013,nadj2013proposal} in contact with conventional superconductors.
For the sake of concreteness, in the present work we concentrate on hybrid semiconductor-superconductor devices as a particularly promising experimental platform for scalable quantum computation based on MZMs~\cite{karzig2017scalable}, although similar ideas to the ones described below could be profitably applied in other systems as well.

An experimental signature of MZMs is the occurrence of a zero bias peak (ZBP) in the tunneling conductance of a normal probe contacting the expected location of a MZM, such as the end of a nanowire~\cite{bolech2007observing,law2009majorana}.
Experiments on semiconductor-superconductor systems have focused on this signature because of its simplicity and practical feasibility.
Several ground-breaking experiments~\cite{mourik2012signatures,das2012zerobias,deng2012anomalous,churchill2013,deng2016majorana,suominen2017zero,nichele2017scaling,fornieri2019evidence,vaitiekenas2020fullshell} observed the predicted signature in a variety of material systems, but also revealed important puzzles and obstacles towards the unambiguous detection of the topological phase.
The measurements did not clearly observe the occurrence of a topological transition accompanied by a gap closing before the onset of ZBPs, suggesting that a non-topological explanation of the ZBP was still possible.
Furthermore, there was often a ``soft" gap profile in the tunneling conductance as a function of bias voltage, which revealed the presence of a substantial subgap density of states, whose presence is detrimental for topological quantum computing based on MZMs.
In addition, it was initially not possible to controllably tune the coupling of the leads to the putative MZM so that the tunneling conductance had the predicted bias voltage and temperature dependence.

Since the first reports of ZBPs, material improvements have rectified some of these problems~\cite{krogstrup2015epitaxy,gazibegovic2017epitaxy,zhang2017ballistic, khan2020transparent, pendharkar2019, kanne2020epitaxial}.
In particular, the advent of epitaxial semiconductor-superconductor interfaces has improved the quality of induced superconductivity and mitigated the presence of subgap states, resulting in the observation of a ``hard'' superconducting gap in tunneling conductance experiments~\cite{chang2015hard,gul2017hard}.
The tunability of the ZBP as a function of coupling to the normal lead and temperature has also been reported~\cite{nichele2017scaling}.

In parallel to these experimental advances, local signatures of MZMs received theoretical scrutiny as it was gradually realized that disorder~\cite{brouwer2011topological,liu2012zero,bagrets2012classD,pikulin2012zero,sau2013density,mi2014x,woods2019zeroenergy,pan2020physical} and potential variations at the ends of a nanowire~\cite{kells2012near,liu2017andreev,reeg2018zeroenergy,vuik2019,stanescu2019robust} can induce ZBPs of trivial origin.
These ZBPs can mimic many of the features of MZMs, including quantized peak heights associated with trivial Andreev bound states (ABSs) induced by smooth potentials at the end of a nanowire~\cite{moore2018,vuik2019}, and some degree of stability with respect to tuning parameters in either case.
Trivial ABSs have been observed in devices with equivalent materials and geometry to those used in the search for MZMs~\cite{lee2014spin,chen2019ubiquitous,valentini2020nontopological}.
The presence of these alternative scenarios essentially prevents the unambiguous identification of MZMs solely via local tunneling conductance measurements. This difficulty is illustrated for instance by the recent retraction of the claim of MZMs based on measurements of quantized ZBPs as detailed in Refs~\cite{zhang2021retraction,zhang2021large}.

Alternative methods to detect MZMs have already been investigated, often based on the $4\pi$-periodic Josephson effect predicted to occur when two topological superconductors are coupled via a junction~\cite{kitaev2001unpaired,kwon2004fractional,fu2009josephson}.
Signatures consistent with the $4\pi$-periodic Josephson effect have been observed soon after the first ZBP experiments~\cite{rokhinson2012fractional}, and have also been seen in a proximitized topological insulator~\cite{wiedenmann2016four} and in the emission spectrum of semiconductor-superconductor junctions~\cite{laroche2019observation}.
Finally, photon-assisted tunneling consistent with the coherent hybridization of MZMs was also observed~\cite{van2020photon}.
However, such microwave signatures have not been consistently detected in systems showing ZBPs~\cite{sabonis2020destructive} and, since they rely on a local coupling between MZMs, they are equally susceptible to the problem of distinguishing MZMs from trivial ABSs.
Thus, these experiments are not suited either to reliably tune into a topological phase.

The difficulties described thus far highlight the need for a more robust way of identifying the topological regime. A key step towards this goal is to widen the scope from finding specific promising points in the parameter space that are consistent with a Majorana interpretation, to instead focus on establishing the presence of a stable topological phase and the accompanying phase transition. The latter manifests as a bulk gap closing and reopening. Combining this global property of a device with local information is much less likely to be mimicked by non-topological effects. This suggests that incorporating a non-local measurement in the search for MZMs is a promising way to tighten the selection criteria and improve the reliability of the identification of topological phases.

Non-local transport through a proximitized nanowire can be accessed in Coulomb blockade spectroscopy experiments~\cite{albrecht2016exponential,vaitiekenas2018selective,ofarrell2018hybridization,shen2018parity,vaitiekenas2020fullshell}, where the superconductor is floating.
An advantage of this approach is that it allows a finer energy resolution than tunneling conductance measurements, potentially enabling the measurement of the energy splitting due to the finite-size coupling between MZMs at the end of a wire.
Combining measurements on wires of different length, this would allow to not only probe the localized nature of candidate MZMs but also the quantitative extraction of the coherence length in the topological phase.
However, this method is also not immune to the difficulty associated with ruling out coincidental trivial ABSs~\cite{chiu2017,shen2020parity}.
Furthermore, a measurement of the bulk energy gap in a Coulomb blockaded device requires tuning the device to a charge degeneracy point to eliminate Coulomb effects, making a systematic attempt rather involved due to the very fine gate voltage resolution required.

\begin{figure}[t]
	\includegraphics[width=\columnwidth]{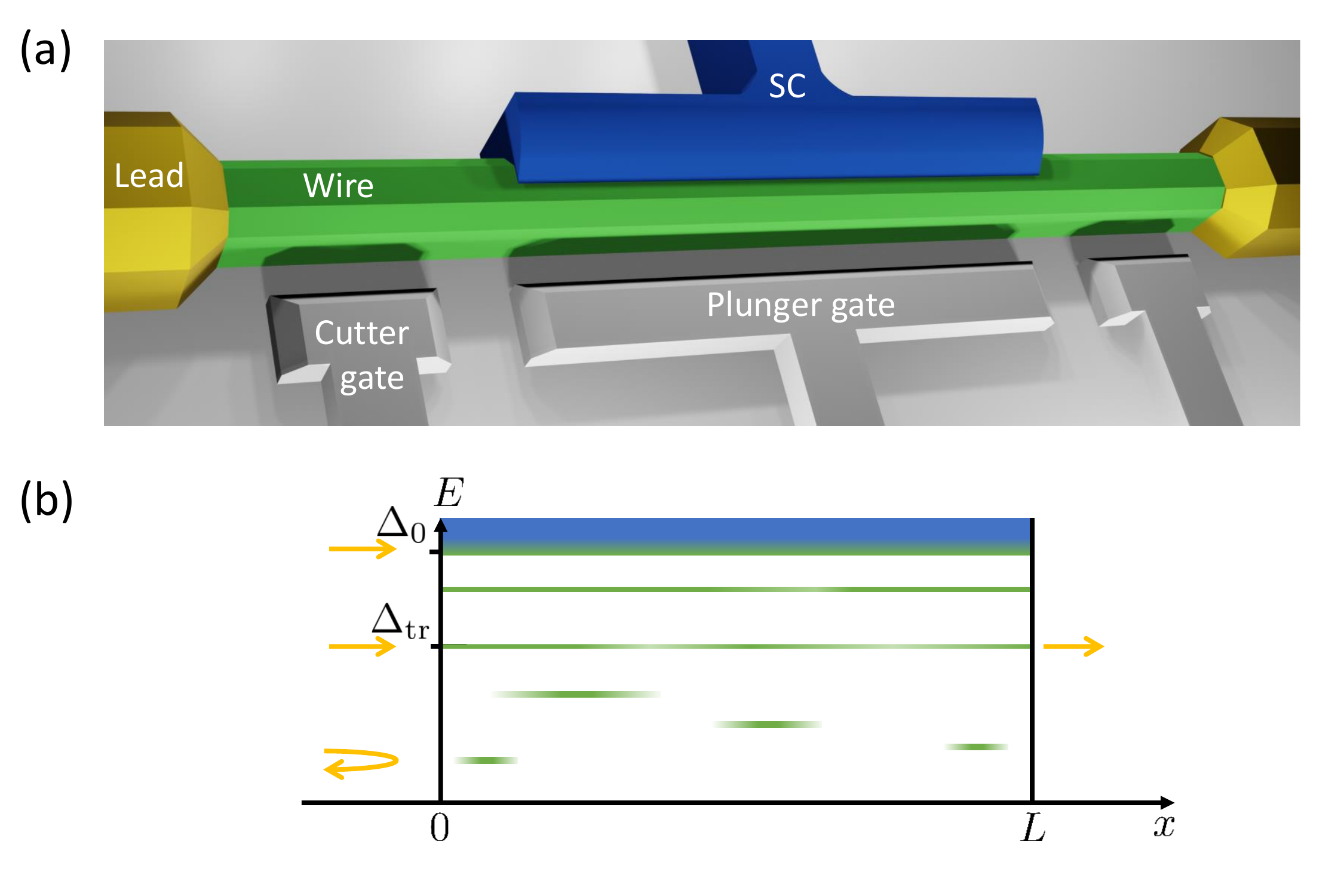}
	\caption{\textbf{Setup and schematic transport through the system.} (a) Schematic of the three-terminal device showing the wire (green) with normal (gold) and superconducting (SC, blue) contacts and plunger and cutter gates (grey).
	The plunger gate is tuning chemical potential in the bulk of the wire and cutter gates are controlling tunneling to the leads.
	(b) Cartoon of the states in the system and their contribution to transport.
	The wire hosts localized states which may show up as peaks in the local conductance.
	At the energy of the transport gap $\Delta_{\rm tr}$ there is the first delocalized state connecting the two sides of the system which is seen in non-local conductance.
	At energies above the parent superconducting gap $\Delta_0$ the spectrum becomes continuous and the particle can escape into the superconducting lead.
	The processes shown may have either normal or Andreev character.
	}
	\label{fig:design}
\end{figure}

An alternative proposal~\cite{rosdahl2018andreev, danon2020nonlocal, pan2021three} to probe the bulk gap of a nanowire is via non-local conductance measurements (also called ``cross-conductance'' in the literature) in three-terminal devices~\cite{menard2020conductance, puglia2020closing} consisting of a proximitized middle segment, which is grounded through a trivial superconductor, and two normal leads on the two sides, see Fig.~\ref{fig:design}.
Such a device geometry enables the simultaneous measurement of local conductance from the two normal leads to ground, thereby allowing us to correlate zero bias features at the two ends of the proximitized nanowire~\cite{anselmetti2019end-to-end}.
Additionally, the non-local signal between the two normal terminals contains information on the energy gap in the entire semiconductor-superconductor segment.
Namely, in a wire sufficiently long compared to the superconducting coherence length, the onset of the non-local signal occurs at bias values corresponding to the lowest energy of an extended mode in the wire~\cite{rosdahl2018andreev}.
This energy is the so-called transport gap as illustrated in Fig.~\ref{fig:design}(b).
Thermal conductance and shot noise power spectrum measurements may be as informative as the electrical conductance for the purpose of the search for the transition~\cite{akhmerov2011, pan2021three, denisov2021charge}.
However, these measurements are more difficult to perform than the non-local conductance and for the purpose of the present manuscript we will not consider them.

We believe that a systematic measurement of the full conductance matrix of a three-terminal device is a promising way to tackle all the aforementioned problems.
In this paper, we formalize and propose a concrete experimental protocol (Topological Gap Protocol or TGP from now on) to identify topological phases based on local \emph{and} non-local measurements in three-terminal devices.
The TGP is a predetermined experimental procedure that yields a yes/no answer to the question of whether a putative topological phase has been identified, the typical value of the topological gap, and the parameter regions in the space of gate voltages and magnetic field applied to the device where the topological region is present.
The TGP presents much stricter requirements for the device to be identified as a topological superconductor than the presence of a ZBP, and thus increases the likelihood for the identified phases to host genuine MZMs.
As we discuss below, the erroneous identification of a trivial system as topological can be excluded with high confidence by the TGP as it is deliberately designed to reject false positives.
This comes at the expense of possibly rejecting a topological region (false negative):
for instance, a possible scenario involving a ``dark" MZM occurring when the bulk of the system is in the topological phase but the junctions prevent efficient coupling of the leads to the MZMs.
Although such a system is, in principle, in the topological phase, it is not useful without a way to access the MZMs.
Thus, we do not prioritize discriminating this scenario from the one where the system is topologically trivial.
Due to the increased complexity, the TGP is more time-consuming than a simple ZBP search but, as we show in this work, fast measurement techniques and automation help to speed up the run-time.

The remainder of the paper is organized as follows.
In Sec.~\ref{sec:overview} we discuss the basic ideas underlying the TGP and present illustrative examples that highlight the need for its introduction.
In Sec.~\ref{sec:topogap_brief_description} we outline the different steps of the TGP, motivating its division into two main stages.
In Sec.~\ref{sec:phase_one} and~\ref{sec:phase_two} we go over the details of the two stages of the TGP, illustrating the required calibration measurements via a test device as well as TGP outcomes using a simulated dataset.
Finally, Sec.~\ref{sec:conclusions} contains concluding remarks.

\section{Overview of the TGP}
\label{sec:overview}

In this Section we outline the principle ideas that motivated the TGP and how they avoid common false positives that can appear in local measurements.

\subsection{Basic Idea}
\label{sec:design_principles}

One-dimensional superconducting phases are classified by a topological invariant taking the values $\pm 1$.
There are several different formulations of the invariant, which are equivalent in the limit of large systems, where there is a sharp distinction between the phases.
To name a few:
(1) the parity of the number of MZMs at a boundary, which holds equally-well in interacting systems; (2) the sign of the Pfaffian of the matrix that defines the Hamiltonian~\cite{kitaev2001unpaired}, which is only defined if interactions are neglected, and must be computed for periodic boundary conditions; or (3) the topological scattering invariant -- the sign of the determinant of the reflection matrix \cite{akhmerov2011, fulga2011scattering} -- which also requires neglecting interactions but is applied to wires coupled to leads, as in realistic models of experimental conditions.
While the topological invariant is easily computed in simulations, with one or another formulation being more convenient for different choices of boundary conditions, no formulation maps directly onto a readily measured quantity in experiments.
On the other hand, the topological invariant can only change sign when the bulk gap closes
due to a propagating mode through the system: this is the phase transition between trivial and topological phases.
Ironically, the phase transition has a less ambiguous signature than the topological phase itself.

In the present manuscript we concentrate on realistic device geometries and thus use the scattering matrix invariant to characterize the system.
The invariant changes when there is zero-bias particle transport through the system and, thus, the
transport gap is closed.
In a disordered system the transport gap is not necessarily equal to the spectral gap as the spectrum may contain localized states below the transport gap, see Fig.~\ref{fig:design}(a).
It is the transport gap that controls the scattering invariant, even in presence of a continuous density of states~\cite{groth2009theory, akhmerov2011, fulga2011scattering}.
We will call it the gap through the rest of the manuscript.

We can identify phase transition lines in parameter space by measuring the full conductance matrix:
\begin{align}
\label{eq:conductance_matrix}
G = 
    \begin{pmatrix}
        G_{RR} & G_{RL} \\ G_{LR} & G_{LL}
    \end{pmatrix}
    =
  \begin{pmatrix}
        dI_R/dV_R & dI_R/dV_L \\ dI_L/dV_R & dI_L/dV_L
    \end{pmatrix}  .
\end{align}
as a function of the magnetic field and plunger gate voltage which tunes the occupation in the semiconductor, see Fig.~\ref{fig:design}(a).
Here, $I_{L,R}$ are the currents at the left or right lead in Fig.~\ref{fig:design}(a), while $V_{L,R}$ are the voltages on the left or right lead.
As discussed in Ref.~\cite{rosdahl2018andreev}, the off-diagonal terms, $dI_R/dV_L$ and $dI_L/dV_R$ are \textit{non-local conductances} that for sufficiently long systems are only non-zero at bias voltages between the transport gap in the semiconductor and the parent gap in the superconductor (both of which depend on the magnetic field).
When the gap in the semiconductor vanishes, the non-local conductances are non-zero all the way down to zero bias.
This can occur either at a phase transition, or when the gap is small enough that it is less than the temperature, or when the coherence length is longer than the system size.

Consequently, we can map out the locus of phase boundaries in parameter space by measuring the
non-local conductance and using it to track the parameter values at which a gap closes and re-opens.
However, as we will describe in greater detail, the non-local conductance is a relatively
time-consuming measurement while the local conductance can be measured rapidly.
Consequently, our protocol first uses measurements of the local conductances $dI_R/dV_R$ and $dI_L/dV_L$ to identify regions in parameter space where ZBPs are observed at both ends of a wire.
Measurements of the non-local conductances $dI_R/dV_L$ and $dI_L/dV_R$ are then used to search for bulk phase transitions near the boundaries of the regions in parameter space with ZBPs at both ends.
These non-local conductance measurements enable us to trace the phase transition lines that separate gapped superconducting states on both sides of the transition.
In this way, we avoid the difficulty with ZBPs, which is that they may appear as a result of local physics without a bulk gap closing, and also avoid the pitfall associated with looking only for an isolated bulk gap closing, which is that it could, in principle, be an ``accidental'' degenerate gap closing flanked by trivial superconducting states on both sides
of the gapless point.

However, it is important to keep in mind that there will be parameter values at which there are
ZBPs at both ends of the wire even when the bulk of the system is in the trivial phase.
Conversely, the system may be in the topological phase, but varying junction gate voltages may cause the MZMs to couple very weakly to one or both of the leads, resulting in an absence of correlated ZBPs.
Finally, if the electrostatic potential is inhomogeneous, then one part of the system may undergo the transition before the other part.
As a result, there may not be a visible gapless mode throughout the system \footnote{As mentioned above, any sign change of the scattering invariant is accompanied by a zero energy resonance connecting the sides of the system. In an inhomogeneous system, however, this resonance can be only very weakly coupled to the leads due to the local gap in parts of the system. Once temperature broadening is included, the resonance becomes practically invisible.}, thus making the observation of the phase transition very difficult to impossible.
Hence, it is crucial to systematically scan over a range of parameters which are expected to trigger the topological phase transition (chemical potential and magnetic field) and identify gapped regions divided by gap closings.
If correlated ZBPs are present in a large fraction of a region that is separated by a gap closing from the non-topological region connected to zero field, this likely means that the ZBP region is in the topological regime.
Identifying regions of topological phase rather than single ZBP traces also removes possible selection bias and many examples of accidental false-positive features, which we discuss in more detail below.

We note that the TGP does not relax the requirements of the system to be of sufficient quality.
In particular a positive outcome of the protocol requires a material that is clean and homogeneous enough for the observation of the topological phase transition.
The disorder mean free path should be longer than the topological coherence length in the system and long-scale inhomogeneities should be sufficiently weak to support a continuous topological phase throughout the system.
Precise conditions on both of these requirements can be obtained using realistic simulations of the measured devices\cite{antipov2018effects, winkler2019, vaitiekenas2020fullshell, shen2020parity, Kringhoj2021}.

In summary, the systematic observation of a gap closing and reopening which enclose the parameter regions of correlated ZBPs in a three-terminal device can be a strong indication that the device is in the topological regime.
The aim of the TGP is to formalize such a set of measurements and the corresponding data analysis.

\subsection{Illustrative examples of simulated
local and non-local conductances.}
\label{sec:false_negatives}

\begin{figure*}
    \centering
    \includegraphics[width=0.9\textwidth]{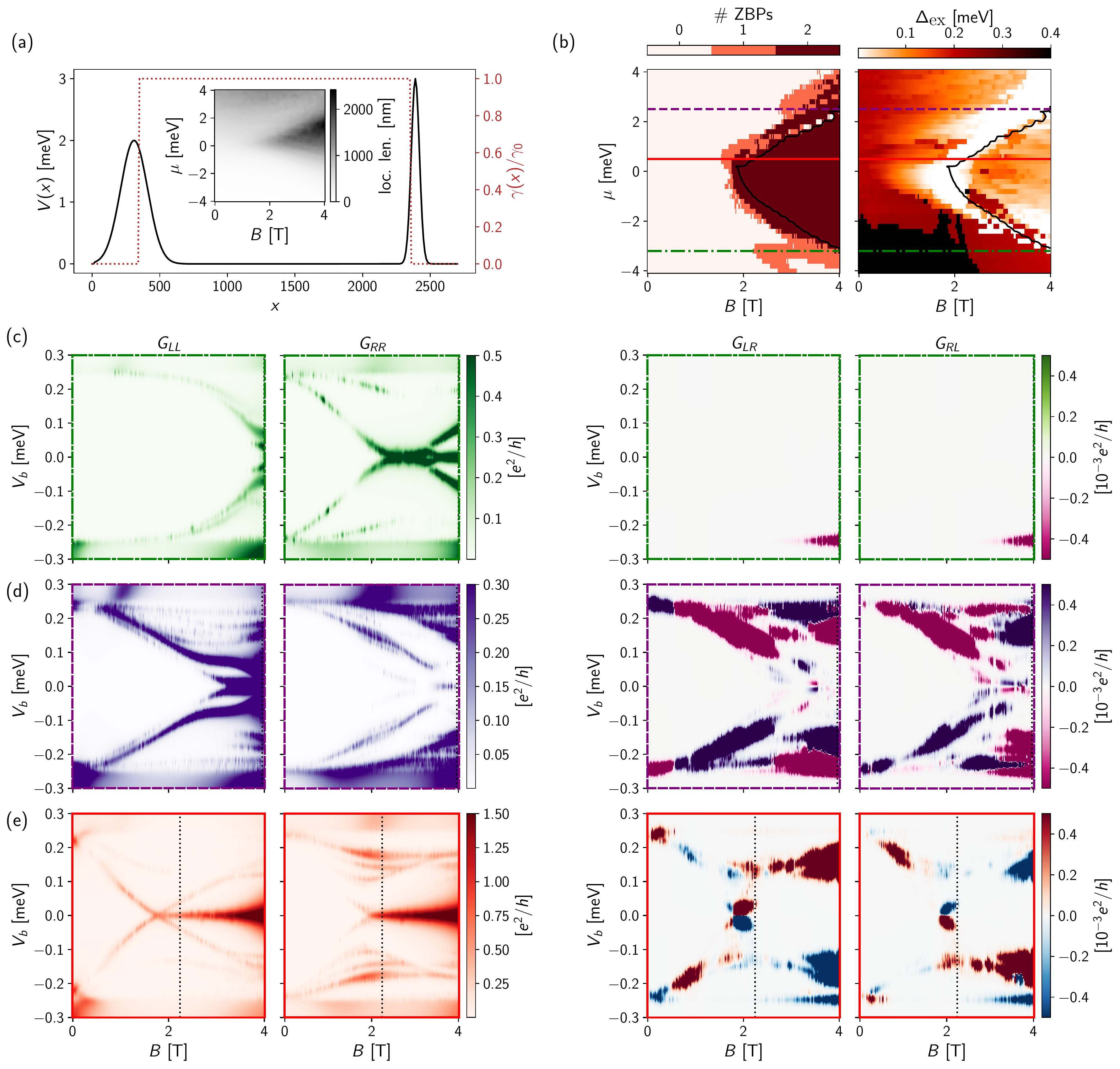}
    \caption{\textbf{Simplified model system with smooth potentials and on-site disorder.} 
    Panel (a) shows the smooth part of the potential $V(x)$ and the spatial dependence of the superconductor-semiconductor coupling $\gamma(x)$.
    The disorder amplitude is fixed and corresponds to normal state localization lengths of $(48, 1435, 845)$ nm for $\mu=(-3.2, 0.5, 2.5)$ meV at $B=2.24$ T, which is the field of minimal clean topological coherence length $128$ nm at $\mu=0$.
    The inset in (a) shows the localization length as a function of $B$ and $\mu$ in the semiconductor.
    (b) shows the identified ZBPs on either side of the system and the extracted bulk transport gap $\Delta_\textrm{ex}$.
    The solid black line indicates a topological phase as identified by a scattering invariant $<-0.9$.
    The three colored horizontal lines indicate the value of $\mu$ for the bias-field scans in (c-e).
    The system exhibits various non-topological ZBPs which can lead to failures of traditional methods of identifying the topological regime, as exemplified by (c) and (d).
    (c) is an example of uncorrelated zero bias peaks without structure in the non-local conductance which indicates a trivially gapped system.
    Row (d) shows correlated zero bias peaks accompanied by a gap closing but no reopening in the non-local conductance.
    All of these are false positives for previous methods but are caught by the TGP.
    Only the example in (e) shows a topological regime which will be correctly identified by the TGP via the appearance of correlated ZBPs and a gap closing and reopening feature.
    The dotted vertical lines indicate the intersection with the black scattering invariant line from (b).}
    \label{fig:false_positive_everything}
\end{figure*}

\subsubsection{Simplified Model and Signatures
of the Topological Phase}

In this subsection, we will consider the simple model of a one-dimensional proximitized Rashba wire~\cite{lutchyn2010majorana, oreg2010helical}.
It is known in the literature~\cite{liu2017andreev, vuik2019, pan2020physical} that this model can give rise to ZBPs that are not due to MZMs. The system consists of a semiconductor (SM) and superconductor (SC) with Hamiltonians (henceforth $\hbar=1$),
\begin{align}
H_{\rm SM} &= \left(\frac{k^2}{2m} - \mu+V\right) \tau_z + \alpha k \sigma_y \tau_z + \frac{1}{2} \mu_B g B 
\sigma_x \tau_z, \hspace{0.5cm}\raisetag{1.5\normalbaselineskip}\label{eq:H_toy1} \\
H_\mathrm{SC} &= \left(\frac{k^2}{2m_\mathrm{SC}} - \mu_\mathrm{SC}\right) \tau_z + \Delta_0 \sigma_y \tau_y, \label{eq:H_toy2}
\end{align}
where $\sigma$ and $\tau$ are Pauli matrices in spin and Nambu space, respectively. The parameters $m$ ($m_{\rm SC}$), $\mu$ ($\mu_{\rm SC}$), $\alpha$, and $g$ denote, respectively, the effective electron mass, the chemical potential, the Rashba spin-orbit strength, and the $g$-factor in the semiconductor (superconductor); $B$ is the applied magnetic field.
A coupling term between $H_{SM}$ and $H_{SC}$ induces a gap in the semiconducting region. Formally this coupling can be described by a self energy \cite{stanescu2011majorana}
\begin{align}
\Sigma_\mathrm{SC}(x,\omega) = \gamma(x) \frac{\omega + \Delta_0 \sigma_y \tau_y}{\sqrt{\Delta_0^2 - \omega^2}},
\end{align}
after integrating out the superconducting degrees of freedom. Here, we introduced the function $\gamma(x)$ which is equal to the coupling parameter $\gamma_0$ for regions of the semiconductor in contact to the superconductor and zero otherwise. We choose parameters suitable to model an InSb/Al hybrid semiconductor-superconductor nanowire \footnote{Here, we neglect the effect of an external magnetic field on the superconductor because its $g$-factor is much smaller than that of the InSb semiconductor and the Zeeman splitting at the magnetic field of interest (i.e.~$B\sim 1$ T) is much smaller than the parent gap.}.
See App.~\ref{sec:details_toy_model_simulation} for the specific parameters and the technical details of the simulation.

The spatially varying electric potential $V=V(x)$ can be used to model tunnel barriers, disorder, and potential gradients.
This simple model is convenient for building intuition on the signatures of the topological phase and of non-topological ZBPs, since the electronic potential $V$ can be chosen specifically to construct known examples of the latter.
In contrast, the more realistic device simulations used in Sec.~\ref{sec:phase_one} and \ref{sec:phase_two}, and described in detail in App.~\ref{sec:details_realistic_simulation}, derive the potential self-consistently from the geometry and electrostatics of the device, which makes the targeted creation of counter examples much more involved.

Fig.~\ref{fig:false_positive_everything} shows some instructive local and non-local conductance plots occurring in the simplified model.
Fig.~\ref{fig:false_positive_everything}(a) depicts the smooth part of the potential profile that is implemented on top of a constant chemical potential $\mu$ and short range disorder in $H_{\rm SM}$ with fixed amplitude.
The disorder strength is depicted by showing the normal state localization length -- obtained using the transfer matrix approach~\cite{DeGottardi_2011} --  as a function of the magnetic field and the chemical potential.
For fixed disorder strength, the system varies between the weakly and strongly disordered regime depending on the magnetic field and chemical potential.
The inset in Fig.~\ref{fig:false_positive_everything}(a) shows a gapped insulator phase at large negative $\mu$ and a regime with longer localization lengths within the Zeeman gap, where back-scattering is suppressed due to spin-orbit coupling~\cite{brouwer2011topological}.

Fig.~\ref{fig:false_positive_everything}(b) shows the number of ZBPs extracted from the local conductance and the extracted gap $\Delta_\textrm{ex}$ from the non-local conductance throughout the phase diagram.
Here, \#ZBPs = 1 when there is a ZBP at one end of the wire but not at the other; \#ZBPs = 2 when there are ZBPs at both ends of the wire.
Details of the corresponding data analysis are discussed in Sec.~\ref{sec:phase_one} and \ref{sec:phase_two}.
The topological phase as extracted from the scattering invariant of the system is marked by the black contour line in (b).
The rows of panels (c), (d) and (e) show the conductance matrix for cuts at fixed chemical potential $\mu$ through the phase diagram.
The cuts are chosen to exemplify different scenarios of interest.

\begin{figure*}[th]
    \centering
    \includegraphics[width=0.95\textwidth]{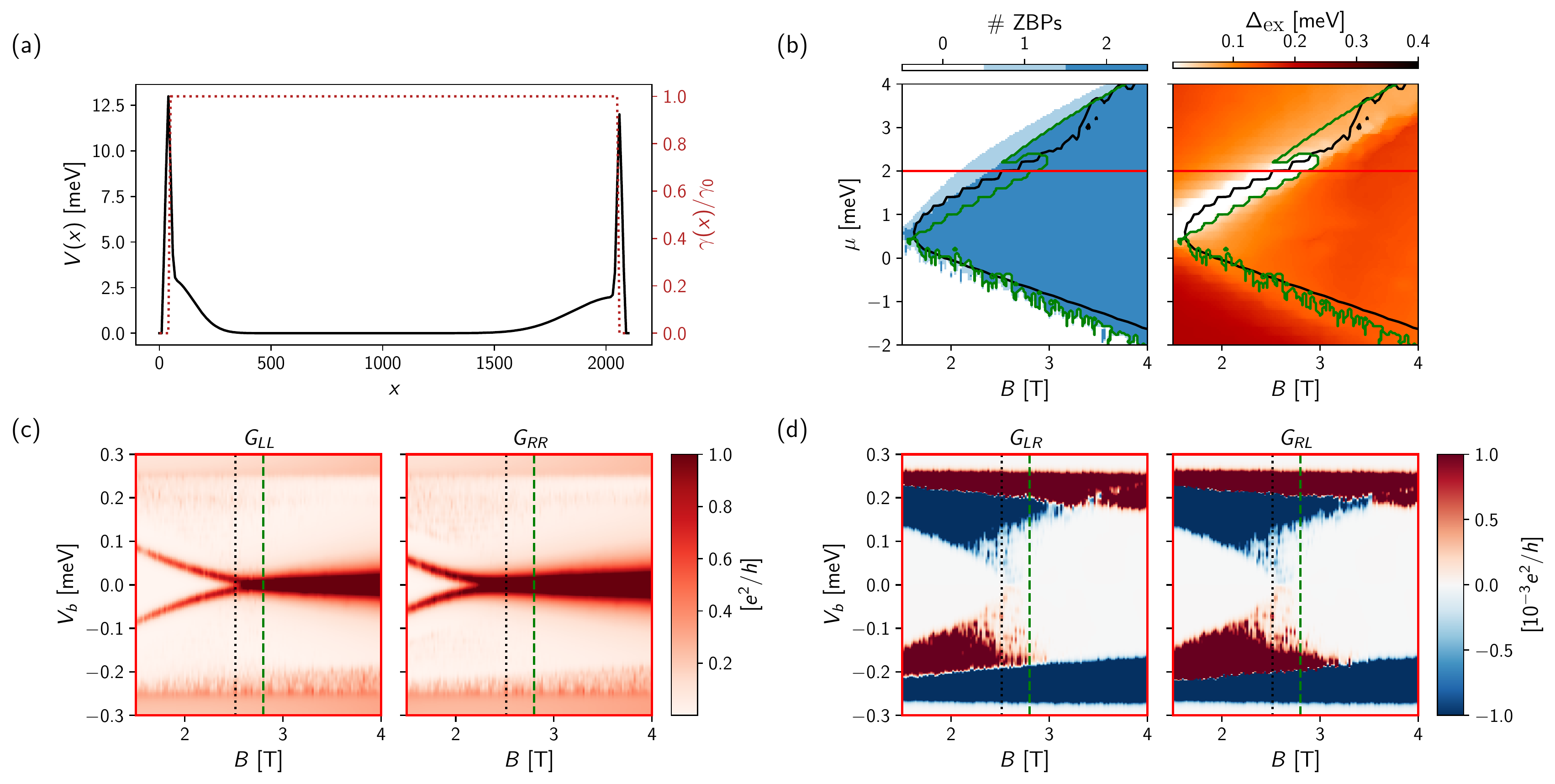}
    \caption{\textbf{Simplified model phase diagram for quasi-Majorana scenario.}
    Potential and superconductor-semiconductor coupling shown in (a), phase diagrams obtained from ZBP extraction and gap extraction from conductance matrix in (b).
    The solid black line indicates a topological phase as identified by a scattering invariant $<-0.9$, while the green line indicates a region with ZBPs on both ends and a gap $>\SI{20}{\micro\eV}$ (corresponding to ROI$_2$, see Sec.~\ref{sec:topogap_brief_description}).
    The median gap of the green region is $\SI{134}{\micro\eV}$ with $30\%$ of its boundary determined to be gapless.
    Local (c), and non-local (d) conductance examples.
    The vertical black dotted and green dashed lines in (c) and (d) indicate the intersection with the lines of corresponding color in (b).}
    \label{fig:app_2QM}
\end{figure*}

The effect of disorder is particularly strong at the bottom of the band ($\mu$ below the Rashba band crossing for $B=0$) which can lead to disorder-induced ZBPs, as in \ref{fig:false_positive_everything}(c), which corresponds to the lower line cut in Fig.~\ref{fig:false_positive_everything}(b).
Since they are due to local effects, their appearance at the two ends of the system is uncorrelated and the one shown in the Fig.~\ref{fig:false_positive_everything}(b) and (c) are specific to the disorder realization.
We also note that the low-energy non-local conductance is featureless around the appearance of the local ZBP.

Fig.~\ref{fig:false_positive_everything}(d) corresponds to the upper line cut in Fig.~\ref{fig:false_positive_everything}(b).
It illustrates another unwanted scenario for the purpose of identifying MZMs that are of interest for topological quantum computation: gapless nanowires that show ZBPs on both ends of the wire simultaneously.
We see that while the local conductance identifies correlated ZBPs, the finite non-local conductance close to zero bias does not indicate the presence of a gapped phase after the appearance of the ZBPs.
This emphasizes the importance of combining local and non-local conductance information.

Finally, Fig.~\ref{fig:false_positive_everything}(e) corresponds to the middle line cut in
Fig.~\ref{fig:false_positive_everything}(b).
It shows the appearance of topologically protected MZMs, where the disorder acts perturbatively with the normal state localization length (\SI{1.4}{\micro m}) being much longer than the minimal clean coherence length (\SI{128}{\nano m}) in the topological region.
Here one can identify the correlated appearance of ZBPs at both ends of the system in conjunction with a gap closing and reopening feature in the non-local conductance.

\subsubsection{Smooth potential variations and quasi-Majorana modes}
\label{subsec:smooth_side}

Variations in the potential near the ends of the wire that are smooth on the scale of the Fermi wavelength can create regions housing low-energy states in magnetic field.
The corresponding low energy modes are often referred to as quasi-Majorana modes.
They are somewhat stable in parameter space given a sufficient potential smoothness and can mimic many of local MZM characteristics~\cite{vuik2019}.
They, however, do not actually enjoy the topological protection required for high-fidelity quantum computation~\cite{mishmash2020}.
Therefore, they may produce false positive results in the conventional ZBP search.

A typical scenario where quasi-Majorana modes appear is when the system is tuned close to but outside of the topological phase.
For concreteness, consider an example where at a fixed magnetic field the chemical potential $\mu$ is larger than the critical chemical potential $\mu_{\rm C}$ required to enter the topological phase.
A smooth potential variation can be interpreted as a spatially varying chemical potential $\mu_{\rm eff}(x)=\mu -V(x)$.
In the above scenario, it is possible for a potential bump close to the end of the wire to locally tune the system into the topological regime $\mu_{\rm eff}(x)<\mu_{\rm C}$, which leads to a local pair of Majorana modes.
The latter manifest in the local conductance as zero bias peaks at lower fields than required for the topological phase transition in the bulk of the wire.
We perform a numerical conductance simulation corresponding to the potential landscape depicted in Fig.~\ref{fig:app_2QM}(a) which includes tunneling barriers and potential variations of different smoothness at each end of the system.
The quasi-Majoranas formed in such a potential are exemplified in Fig.~\ref{fig:app_2QM}(b) where for positive $\mu$ the zero bias peaks appear at lower fields than the critical field of the topological transition.
The latter is denoted by the black contour line in the Fig.~\ref{fig:app_2QM}(b) which is extracted from the scattering topological invariant.
The smoother the potential variation, the earlier in magnetic field a zero bias peak emerges, which can be seen in the local conductance plots in Fig.~\ref{fig:app_2QM}(c).

The effect of smooth potentials on the non-local conductance is more nuanced.
The non-local conductance should still detect the closing and reopening of the bulk gap.
The smooth potential, however, can lead to a reduced visibility of the bulk phase transition, see Fig.~\ref{fig:app_2QM}(d).
Any variation from the bulk value of the potential at the phase transition $\mu_{\rm eff}(x)=\mu_{\rm C}$ leads to a gapped region which suppresses the low bias non-local conductance and reduces the visibility of the gap closing.
This effect can be quantified by the gapless fraction of the boundary of the region that is gapped and exhibits ZBPs at both ends, see green contour line in Fig.~\ref{fig:app_2QM}(b).
For the particular example only 30\% of the boundary is determined to be gapless (see Sec.~\ref{sec:phase_two} for details) which would suggest a false-negative label of the region.

Thus, a smooth potential could cause us to look for a gap closing at $B$ fields that are too low while also obscuring the gap closure that does occur at higher fields.
Finally, at negative $\mu$ in the simplified model, it is also possible for the reverse situation to occur: the bulk of the system is in the topological phase, but the smooth potential $V(x)$ creates regions of trivial phase near the two ends which act as large barriers between the MZMs and the leads, preventing the observation of ZBPs.

\subsubsection{Disorder}
\label{subsec:disorder}

\begin{figure*}[th]
    \centering
    \includegraphics[width=0.95\textwidth]{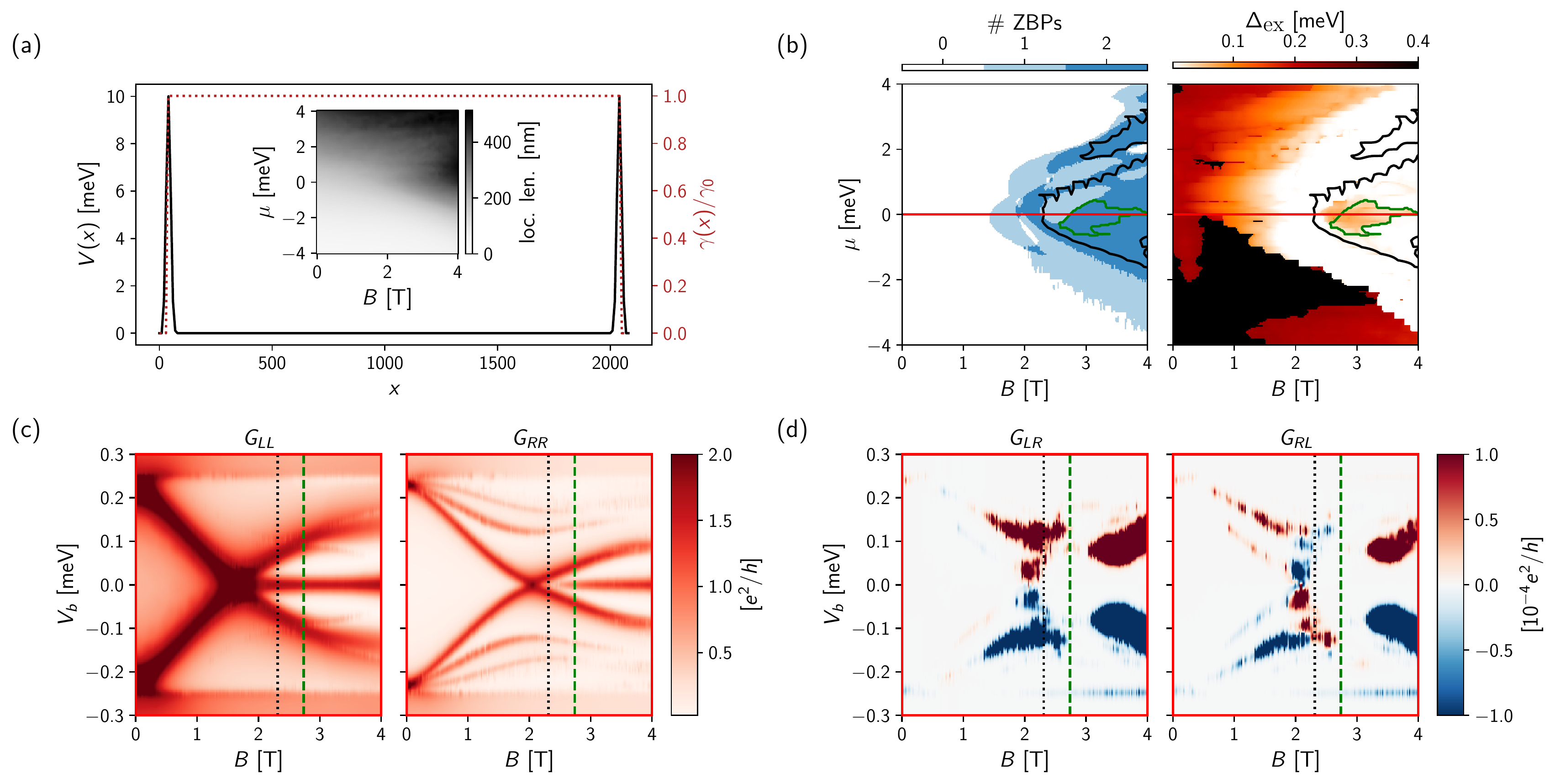}
    \caption{ \textbf{Simplified model phase diagram for moderately strong disorder}.
    Superconductor-semiconductor coupling and smooth part of the potential shown in (a).
    The inset shows the normal state localization length of the semiconductor as a function of $B$ and $\mu$.
    For comparison the minimal topological coherence length in the clean system is $128$ nm at $\mu=0$ and $B=2.24$ T while the localization length from inset (a) at that point is $243$ nm.
    Phase diagrams obtained from ZBP extraction and gap extraction from the conductance matrix in (b).
    The solid black line indicates a topological phase as identified by a scattering invariant $<-0.9$, while the green line indicates a region with ZBPs on both ends and a gap $>\SI{20}{\micro\eV}$ (corresponding to ROI$_2$, see Sec.~\ref{sec:topogap_brief_description}).
    The median gap of the green region is $\SI{41}{\micro\eV}$ with $93\%$ of its boundary determined to be gapless.
    Local (c) and non-local (d) conductance examples.
    The vertical black dotted and green dashed lines in (c) and (d) indicate the intersection with the lines of corresponding color in (b).
    \label{fig:app_moddisorder}}
\end{figure*}

As established by previous studies~\cite{Motrunich_2001, brouwer2011topological,liu2012zero, Lobos_2012, ioselevich2012majorana,bagrets2012classD,pikulin2012zero,sau2013density,mi2014x,woods2019zeroenergy,pan2020physical}, disorder may lead to the appearance of trivial ZBPs and may even destroy the topological superconducting phase.
A quantum phase transition between the topological superconducting phase and the disordered phase (i.e.~an Anderson insulator) occurs when the clean topological superconducting coherence length $\xi_P$ is equal to the normal-state localization length $l_c$~\cite{Motrunich_2001, brouwer2011topological, Lobos_2012}.
It can be quite challenging to distinguish genuine MZMs and disorder-induced trivial ZBPs.
Fig.~\ref{fig:app_moddisorder} shows simulation results for a system with a stronger disorder than in Fig.~\ref{fig:false_positive_everything}(c).
The disorder strength is quantified by the normal state localization length in the inset of Fig.~\ref{fig:app_moddisorder}(a), which becomes shorter than the topological coherence length in some parts of the phase diagram while still leaving a smaller region of topological phase.
A common (but not necessary) characteristic of disorder-induced ZBPs, noticeable in Fig.~\ref{fig:app_moddisorder}(b), is that they show up at one side of the system since they arise from local fluctuations in system parameters.
These peaks may be stable to some degree of field and chemical potential variation but there is a very small region of the phase diagram in which there are ZBPs at both ends of the wire.

Disorder can soften the gap~\cite{brouwer2011probability, lutchyn2012momentum, takei2013soft}, suppressing the topological phase.
This effect is apparent by the increase in the low gap (white) regions in Fig.~\ref{fig:app_moddisorder}(b).
Importantly, while some of the gapless regions technically qualify as topological from the perspective of the scattering invariant, they are not useful for topological quantum computation, because a significant density of subgap states present throughout the device causes quick decoherence of the MZMs~\cite{karzig2021quasiparticle}.
Requiring both the presence of correlated ZBPs and a finite gap leads to a smaller region marked by the green contour line in Fig.~\ref{fig:app_moddisorder}(b).
As described below, the TGP is designed to find such regions with potential use for quantum computation.

A potentially more serious problem is that disorder reduces the strength of the non-local signal since impurity scattering reduces the chance of above-gap quasiparticles to propagate from one end to the other.
Such a signal reduction is hinted at in Fig.~\ref{fig:app_moddisorder}(d) and can make the gap extraction procedure described below more challenging.
In the extreme limit, the signal may becomes too weak for signatures of a gap closing and reopening to be visible at all.
For local conductances of $1 e^2/h$ as shown in Fig.~\ref{fig:app_moddisorder}(c), and assuming the effective voltage noise in the experimental setup to be around \SI{10}{\nano\volt}, the expected noise floor for the non-local conductance measurements is on the order of $10^{-3} e^2/h$, which could render signals as small as those in Fig.~\ref{fig:app_moddisorder}(d) undetectable.

\subsubsection{Long-ranged potential variations along the wire}

Long-range potential variation along the wire can make the topological transition in parameter space non-simultaneous in different regions of the wire.
This would still allow for a topologically non-trivial region in parameter space, but it would not be surrounded by a bulk gap closing.
A prime example of this effect is the presence of a potential gradient in the system, which could be due to
misalignment of the wire with the plunger gate.
Weak gradients will lead to a reduced visibility of the phase transition in the non-local conductance, see Fig.~\ref{fig:app_gradient} in the Appendix.
For stronger gradients the phase transition might no longer be resolved for practical values of the non-local conductance.
This situation will lead to a clear false negative of the TGP that we discuss in App.~\ref{sec:examples_rashba_wire}.
This risk should be mitigated via careful device design outside of the execution of the TGP.

\section{Description of the TGP}
\label{sec:topogap_brief_description}

Based on the general principles of Sec.~\ref{sec:design_principles} and with the possible scenarios just described in mind, we have formulated an experimental prescription for the TGP which we outline here and describe in detail in the next two Sections.

\subsection{Practical criteria}

The parameter space of a typical device aimed at the detection and manipulation of MZMs is very large -- at least three gates and the magnetic field control the state of the system and should all be scanned to search for the topological parameter regime.
Furthermore, the topological regions of the parameter space may be small compared to the total range of a given control parameter that is accessible experimentally, especially in the case of gate voltages (see for instance the topological phase diagrams reported in Ref.~\cite{winkler2019}).
This requires scanning the parameter space with a fine resolution in order to make sure that all interesting regions can be captured.
Because traditional lock-in measurements of the conductance can take $0.1 - 0.5$s per data point, a naive scan of the parameter space with a fine enough resolution can easily take weeks or longer.
These long measurement times are problematic also in view of the fact that the operational stability of the device may have shorter time scales, e.g.~due to uncontrollable charge re-configurations in the dielectrics.

For this reason, MZM search experiments have often resorted to heuristic methods, for instance via the execution of a finite sequence of one-dimensional scans through the large parameter space, chosen by intuition, experience, and/or theoretical considerations to have the highest chance of finding a ZBP.
While these methods can be quick compared to an extensive scan, they are subject to selection bias due to several reasons: the region explored may not be representative of the behavior of the device; the sequence and total number of measurements are not predetermined; and ZBPs may have multiple origins, not necessarily related to a topological phase.
After a ZBP is found, the likelihood for it to correspond to a genuine MZM is often tested with additional measurements which rely on a notion of stability, i.e.~on the requirement that the ZBP persists in a finite range in gate voltages and field.
However, while such an analysis can help rule out some of the most obvious false positives, operationally it has proven hard to define and quantify a useful notion of stability.
This is because the expected range of stability of a genuine MZM as a function of different tunable parameters depends on the material properties as well as on the lever arms of the gates.
These properties are often, if not always, unknown \emph{a priori} for a given device, and possibly subject to sample-to-sample fluctuations.
Thus, the only meaningful notion of stability is to require a ZBP to persist as long as the bulk gap of the system does not close which, as previously discussed, requires a complementary non-local measurement.

A way to overcome the challenge of the large parameter space is to adopt a faster measurement technique than traditional lock-in conductance measurements.
A valid alternative is the adoption of radio frequency (RF) reflectometry techniques, which have been proven to be a quantitatively accurate surrogate for local conductance measurements~\cite{harabula2017measuring,razmadze2019radio}, with speed-ups of several orders of magnitude in measurement time.
As we show in more detail below, the adoption of these techniques makes it possible to perform an exhaustive parameter space scan of the local conductances of a three-terminal device.

\subsection{Stages of the TGP}

The circumstances just described necessitate the separation of our Protocol into two stages: a first stage based on the measurements of the local signals at RF frequencies, and a second stage based on the measurement of the full conductance matrix at near-DC frequencies.
This separation narrows down the region of parameter space where the slower non-local conductance measurements are performed.

To be more specific, the steps of the TGP are:
\begin{enumerate}
	\item Stage one:
	\begin{enumerate}
		\item Perform calibration between RF and DC conductance measurements.
		\item Perform an RF reflectometry measurement of the local conductances on the left and on the right sides of the device as a function of the bias, the corresponding cutter gate voltages, plunger gate voltage, and magnetic field.
		Map the measured RF signals to local conductances using the calibration data.
		\item Find the positions at which zero bias peaks occur in the local conductances.
		\item Group the zero bias peaks into regions and check if there are regions of interest (ROI$_1$) where ZBPs are present on both sides of the wire.
		This step already cuts off some of the false positives discussed previously as disorder only accidentally causes simultaneous ZBPs on both sides of the wire~\cite{lai2019presence}.
	\end{enumerate}
	The output of Stage 1 is a sequence of ROI$_1$s specified by ranges of gate voltages and field.
	\item Stage two:
	\begin{enumerate}
		\item Perform DC measurement of the full 2$\times$2 conductance matrix~\eqref{eq:conductance_matrix} for each of the ROI$_1$.
		\item Estimate the transport gap of the system by determining the onset of the non-local conductance signal with respect to the bias voltage.
		\item Search for a region of parameter space that has ZBPs measured on both ends of the wire and a finite gap.
		Additionally check that the boundary of that region is gapless.
		This step ensures that the region is consistent with an extended topological phase bounded by a gapless region in parameter space, and thus further removes possible false positives like trivial ZBPs that do not produce a gap closing.
		The regions satisfying these constraints will in general be a subset of the ROI$_1$, which we will denote as ROI$_2$.
	\end{enumerate}
\end{enumerate}

While the realization of the TGP just described above is a way to meet all of its desired requirements, it is not the only possible way.
We focused on scanning gate voltages and magnetic field, but alternative platforms for realizing MZMs may be controlled by different physical parameters, thus affecting the experimental parameter space to be explored to reconstruct a topological phase diagram.
In the next two Sections, we illustrate the execution of the TGP using simulated data in a realistic device geometry as an example.
We will execute the steps of the TGP on this example and briefly describe how the corresponding data can be obtained in experiment.

\section{First stage of the TGP}
\label{sec:phase_one}

In this Section we describe in detail the first stage of the TGP, outlined in Sec.~\ref{sec:topogap_brief_description}.
The principal part of the first stage is the fast measurement of local conductances in RF reflectometry.
The fast local measurement for the three-terminal device is closely related to the fast measurement of a conventional junction, as widely studied in the experimental literature~\cite{reilly2007fast,harabula2017measuring,razmadze2019radio}.
We refer to Fig.~\ref{fig:rf-calibration} for an explanation of the required measurement circuit.
This stage of the TGP is designed with the goal of finding regions of correlated zero bias peaks on the two sides of the device, which allows to quickly characterize a device and identify candidates for topological regions.
We call the regions of interest obtained after the first stage of the protocol ROI$_1$'s.
ROI$_1$'s are then transferred to the second stage, which is described in Sec.~\ref{sec:phase_two}.

In greater detail, the steps of the first stage of the TGP are as follows:
\begin{enumerate}
    \item Perform basic checks on the device.
    These include, for example, the ability to tune both junctions, and finding the magnetic field for which the parent superconducting gap closes.
    We do not go into details here as they depend on the exact experimental system employed.\label{enum:1-1}
    \item Perform RF reflectometry to DC conductance calibration measurements.
    The pinch-off curve of the junction is measured simultaneously with RF reflectometry and with the standard low-frequency lock-in technique.
    This measurement on both sides of the device provides calibration datasets that can be used to map the RF reflectometry signal to the DC conductance [see Sec.~\ref{sec:rf-calibration} for the implementation details].\label{enum:1-2}
    \item Perform a reflectometry measurement of the local RF reflection coefficients on the two sides of the device as a function of magnetic field and plunger gate voltage for a few values of cutter gate voltages.
    The obtained datasets, in combination with the calibration datasets, provide the local conductance values.\label{enum:1-3}
\end{enumerate}

Step~\ref{enum:1-3} is used to execute a systematic RF measurement of the device in the parameter space defined by the plunger gate, the magnetic field, the cutter gates, and the bias voltages.
This parameter space is sampled on a uniform grid, which is chosen according to the following guidelines:
\begin{itemize}
    \item The cutter gates should be set such that both junctions are in the tunneling regime: their high-bias conductance is smaller than $e^2/h$ and it stays within bounds defined as follows.
    The lower bound on the conductance comes from the practical requirement of a visible non-local signal.
    The upper bound is for the system to be in the weakly conducting regime where the local conductance is a convolution of the local density of states with the tunneling matrix element.
    This is the regime where a peak in the conductance signals the presence of a bound state near the junction.
    For our simulations, these bounds mean that the high-bias local conductance is between approximately $0.05~e^2/h$ and $0.25~e^2/h$.
    These values can be translated to a cutter gate voltage range via a measurement of the pinch-off characteristic of each junction.
    \item A magnetic field scan should be performed up to the field at which the bulk gap of the parent superconductor closes.
    If we do not determine and attempt to reach this field, we may be missing the topological regime.
    If the critical field of the parent superconductor is unknown, a practical way to determine it would be to perform a measurement of the conductance matrix as a function of magnetic field and bias, with the plunger gate tuned to the negative limit of its range to minimize the contribution of the semiconductor to the signal by pushing the wave functions into the superconductor~\cite{antipov2018effects}.
    \item Determine the range of plunger gate voltages that should be scanned.
    This is the dimension of the parameter space on which there is often the most uncertainty with respect to the possible location of topological phases.
    In the absence of any previous information about plunger gate ranges of interest, this range could be simply dictated by purely experimental considerations, e.g.~avoiding breakdown voltages.
    The interesting range could also be restricted with the help of simulations.
    For instance, for InSb/Al hybrids we find from simulations that the following criteria help restricting the plunger gate range: the gate should be scanned from a very negative voltage where no subgap states form at any field value, to the voltage at which the induced gap closes at $\approx 100$ mT or less, likely signaling the accumulation of electrons away from Al~\cite{shen2020parity}.
    If the experimentally available range is too large to be covered in a single run, it is possible to execute several stage one measurements sequentially.
    \item Determine a grid step in magnetic field and plunger gate voltage small enough to resolve the extent of the topological phase(s) of interest.
    A reasonable criterion is to have at least ten grid points in the maximum extent that a topological phase is expected to have along both dimensions.
    The required resolution should be determined on the basis of simulations of the topological phase diagram of the device, and potentially increased to take into account the uncertainty on the lever arm of the plunger gate and on the $g$-factor.
    \item Determine a grid step in cutter gate to have at least five data points in the previously specified $0.05-0.25~e^2/h$ high-bias conductance range.
    With this minimum requirement, local effects due to fine-tuning of the operating point of the tunnel barriers can be excluded.
    If the total run-time of the measurement allows, the resolution can of course be increased.
    \item It is not needed to make the bias voltage step finer than the energy resolution set by temperature, coupling to the leads, and noise broadening.
    We suggest using steps of $\leq 5-\SI{10}{\micro\volt}$, which are comparable to the experimentally accessible electron temperatures of $\approx 50$ mK.
    \item Finally, we note that in experiment the plunger and cutter gates can have cross-talk, i.e. changing the plunger gate changes the local conductance at a junction.
    This effect may be compensated for at this stage, for example, by simultaneous tuning of the plunger and cutter gates so that the above-gap conductance remains constant as a function of the combined gate.
	Details of how to do it in practice are beyond the scope of our manuscript.
\end{itemize}

A convenient way to execute the nested loop of the systematic search in parameter space is the following.
The innermost (fastest) loop consists of plunger-bias scans, executed with an optimized gate ramp speed.
The cutter gates are incremented in-between the successive plunger-bias scans, with the magnetic field being changed in the outermost (slowest) loop.
All four combinations of left/right cutter and left/right bias are scanned at each value of the field.
While the bias on each side is scanned, the bias on the opposite side is kept fixed at 0~V, and the cutter on the opposite side is kept at a fixed value which is in the middle of the predetermined range.

Thus, the raw data generated by the first stage of the TGP is the following:
\begin{itemize}
	\item A sequence of calibration datasets consisting of two 2D cutter-field scans, on the left and the right sides, as motivated in more detail later.
	For each point of this scan, three parameters are measured: a trace of the in-phase RF signal, a trace of the out-of-phase RF signal, and the conductance of the respective side.
	\item A measurement dataset consisting of four field-cutter-plunger-bias scans.
	The four scans correspond to the four combinations of left/right bias and left/right cutter.
	The two scans in which cutter and bias are scanned on the same side (both left or both right) are directly used in the data analysis to scan for ZBPs.
	The two additional scans are useful mainly to correct for voltage divider effects~\cite{voltagedivider}, and may be omitted if that is not necessary.
	For each point of this dataset, two parameters are measured: in-phase and out-of-phase components of the RF signal.
\end{itemize}

\subsection{RF circuitry and calibration}
\label{sec:rf-calibration}

\begin{figure}[t]
	\includegraphics[width=\columnwidth]{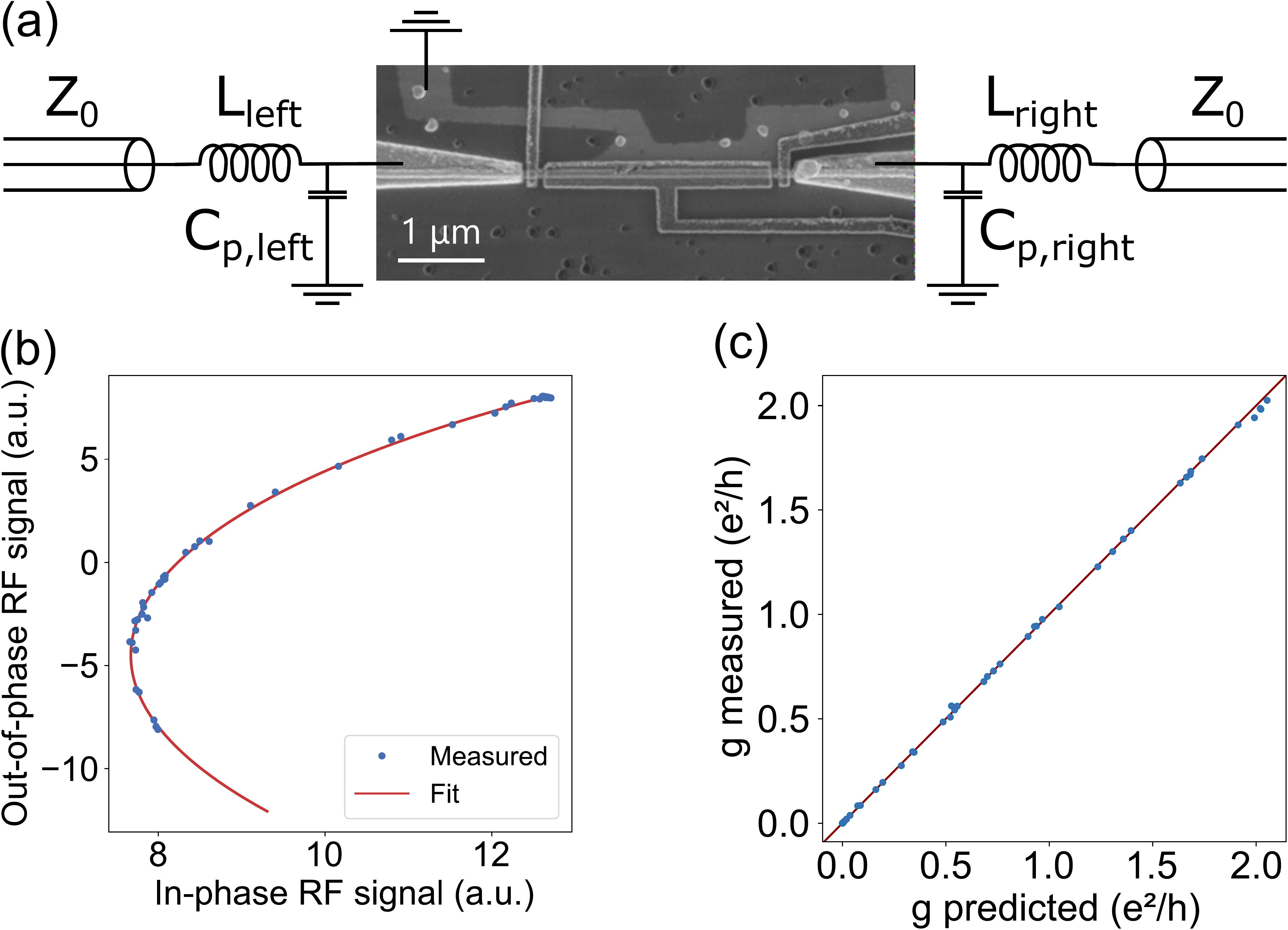}
	\caption{\textbf{RF to DC conductance calibration.}
	(a) Micrograph of test device and measurement circuit diagram.
	(b) Parametric plot of in-phase and out-of-phase components of the reflected RF signal from a calibration dataset on the right side, where the conductance $G_{RR}$ was swept from fully closed ($0 \ e^2/h$) to fully open ($\approx 2 \ e^2/h$).
	The dots are the measured data points, and the solid line is a fit using Eq.~(\ref{eq:rf-transfer-function}).
	(c) Measured DC conductance on a validation dataset vs.~predicted DC conductance, using the model fit from (b).
	The solid line corresponds to the identity.
	\label{fig:rf-calibration}}
\end{figure}

To determine the feasibility of the first stage of the TGP, we demonstrate RF measurements and calibration on the device shown in Fig.~\ref{fig:rf-calibration}(a).
The device was fabricated on a selective-area grown (SAG) InAs nanowire, on which Al was deposited epitaxially.
The Al on the substrate was etched away in order to define a lead contacting the middle of the wire, and ohmic contacts were evaporated on each end of the wire.
After evaporation of a dielectric layer, electrostatic gates were deposited on the ends (cutter gates) and on the bulk (plunger gate) of the wire, according to the design illustrated in Fig.~\ref{fig:design}.
The three-terminal measurement setup and procedure are described in more detail in Refs.~\cite{anselmetti2019end-to-end, menard2020conductance}.
Each of the ohmic contacts was connected through a tank circuit with impedance $L_{\text{left/right}}$ and parasitic capacitance $C_{\text{p, left/right}}$ to a transmission line with characteristic impedance $Z_0$ in order to perform RF reflectometry readout.

We illustrate the calibration procedure on the right side of the device, which was connected to a tank circuit with nominal inductance $L = \SI{150}{\nano\henry}$, estimated $C_p \approx \SI{0.4}{\pico\farad}$, and line impedance $Z_0 = \SI{50}{\ohm}$.
The right cutter gate of the device was swept such that the device conductance spanned the range from fully open ($\approx 2 e^2/h$) to fully closed ($0\,e^2/h$).
For each gate value, the conductance was measured with a lock-in excitation at \SI{94}{\hertz}, together with the reflected RF signal in-phase (I) and out-of-phase (Q) components at $\SI{662}{\mega\hertz}$.

To perform the calibration of the reflected RF signal $S$ to the DC conductance $g$, we assume the sample to be a variable resistor.
The tank circuit is modelled as an impedance $Z_s$ in series with the sample, coming mostly from the inductor $L$, and a parallel impedance $Z_p$ to ground, dominated by the parasitic capacitances.
The reflected RF signal $S$ is related to the incoming RF signal $S_{\text{in}}$ as follows:
\begin{align}
    \label{eq:line-reflection}
    S = S_{\text{in}} \frac{Z - Z_0}{Z + Z_0},
\end{align}
where $Z$ is the equivalent impedance of the tank circuit, given by:
\begin{align}
    \label{eq:equivalent-impedance}
    Z = Z_s + \left(Z_p^{-1} + g\right)^{-1}.
\end{align}
Combining Eqs.~(\ref{eq:line-reflection}) and (\ref{eq:equivalent-impedance}), the transfer function relating $g$ to the reflected complex RF signal $S$ can be written in the form:
\begin{align}
    \label{eq:rf-transfer-function}
    S = S_0\,\frac{ag - 1}{bg + 1},
\end{align}
where $S_0$, $a$ and $b$ are complex coefficients.
The value of these coefficients can be determined empirically: the raw calibration data is plotted parametrically and fitted to Eq.~\eqref{eq:rf-transfer-function} as shown in Fig.~\ref{fig:rf-calibration}(b).
Fig.~\ref{fig:rf-calibration}(c) shows the measured conductance versus the one transformed from RF data from a validation dataset.

We find it necessary to perform this calibration measurement as a function of magnetic field, as the latter can affect the circuit parameters of the tank circuit.
On the other hand, we find that the transfer function is not sensitive to the gate voltages applied to the three-terminal device.
Thus, the transfer function can be used to infer the conductance from all subsequent measurements of the RF signal carried out in the TGP.
Covering a large conductance range in the calibration measurement is required to avoid extrapolation of the transfer function outside of a measured range of conductances.

We estimate that the measurement of a full RF dataset would take on the order of 10 hours.
This assumes our current integration time of \SI{50}{\micro\second} per pixel, which yields acceptable signal to noise ratio on the test device, 200 points for each bias voltage (left and right side), 500 points in plunger gate, 20 cutter gate values, and 200 points in magnetic field.
This timing may be reduced by up to an order of magnitude by improving the impedance matching of the tank circuit to the device and hence the signal-to-noise ratio.

\subsection{Data analysis of Stage 1 measurements}
\label{sec:data_analysis_one}

To illustrate the execution of Stage 1 of the TGP, we performed numerical simulations of the conductance matrix of a three-dimensional model of an InSb/Al proximitized nanowire device in the presence of weak disorder, see details in App.~\ref{sec:details_realistic_simulation}.
We generated a numerical dataset of similar size and resolution as expected from an experimental execution of this stage of the TGP.
Namely, our datasets have a resolution of $\delta V_{\rm plunger} = \SI{1.5}{\milli V}$ in plunger gate and $\delta B = \SI{4.75}{\milli T}$ in magnetic field, covering a total range of 0.6~V and 1.9~T.
Furthermore, we simulate 5 cutter values per side.

We remind the reader that the goal of this data analysis stage is to identify promising regions in parameter space which have a high likelihood to contain an unbroken topological phase.
This occurs via the following steps:

\begin{enumerate}
	\item Convert the RF signal input into conductance, using the transfer function determined with the calibration dataset.
	Since we did not need to model the RF signal separately in simulations, this step was not performed in the current illustration based on the simulated data.
	\item \label{item:classify_bias_traces} Classify each point in the (field, plunger) parameter space as ``promising" or ``not promising",  using as input the $G_{LL}, G_{RR}$ bias traces measured at that point for different cutter voltages.
	The classification first checks for the presence of a ZBP in one of the traces for a given cutter gate voltage as described in App.~\ref{app:data_analysis}1.
	Then, we compute the probability of a ZBP occurring on either side of the device at each (plunger, field) point, simply defined as the fraction of cutter gate values for which a ZBP occurs.
	A (plunger, field) point is classified as promising if both the probabilities are larger than a predetermined threshold.
	In our current example, we chose a threshold of $80\%$, but this number can be adjusted to make the selection more or less stringent.
	\item Find clusters of promising points and filter out those clusters whose volume or shape in parameter space is deemed incompatible with a topological phase.
	In the current example, we use clustering method \texttt{OPTICS}~\cite{scikit-learn}, which is a density-based clustering algorithm that finds collections of promising points which are close together, but do not necessarily form a convex shape.
	Among the obtained clusters, the ones that touch zero field, if any, are filtered out as obviously non-topological.
	The remaining clusters constitute the regions of interest determined by stage 1 of the protocol, ROI$_1$.
	See additional details on the clustering choice in App.~\ref{app:data_analysis}2.
\end{enumerate}

\begin{figure}[t!]
	\includegraphics[width=\columnwidth]{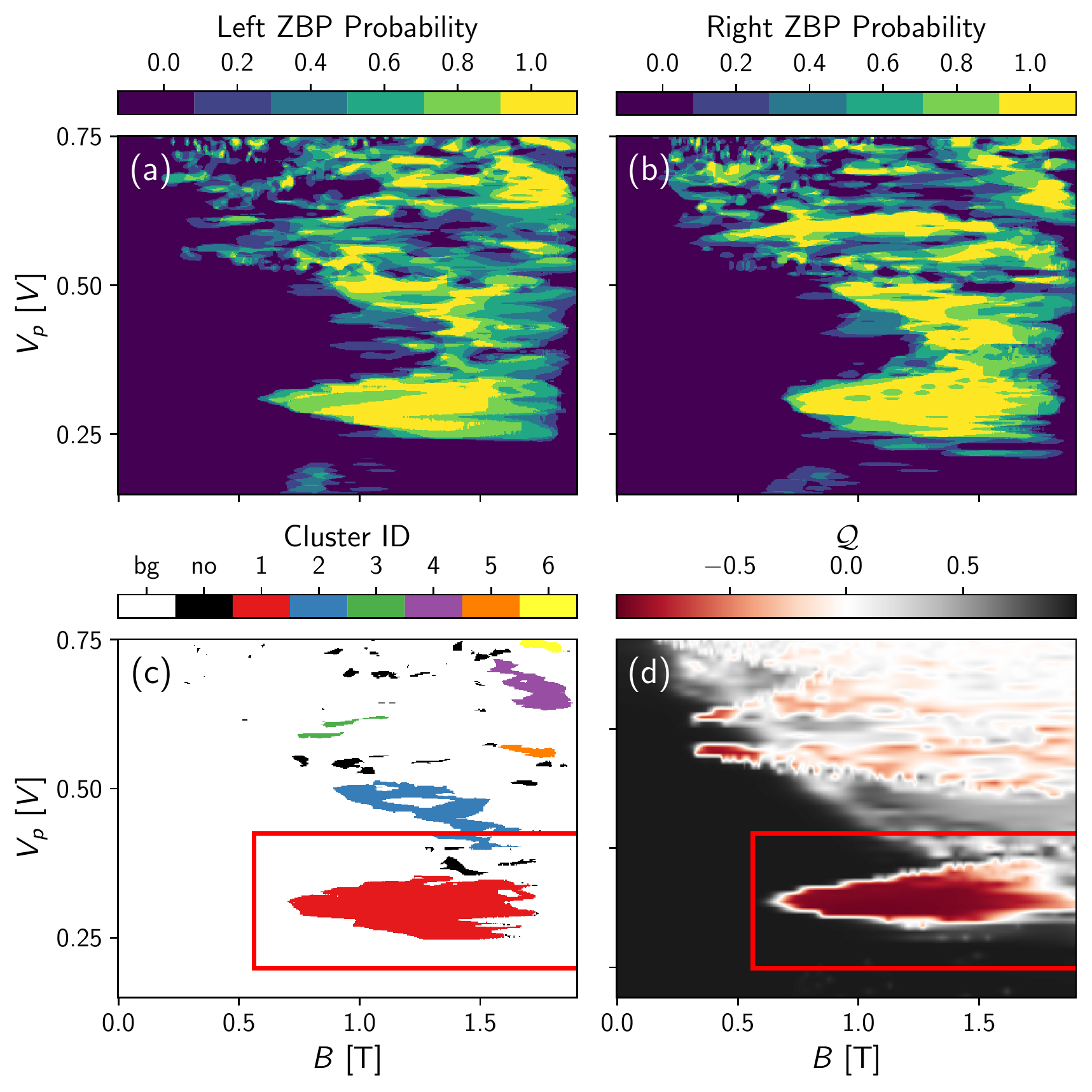}
	\caption{\textbf{Stage one data analysis.} A simulated dataset of an InSb/Al nanowire of  length $L=\SI{3}{\micro\meter}$ is analyzed.
    (a-b): Fraction of cutter voltages for which a zero bias peak appears on the left (a) and right (b) sides of the devices.
    (c): Output of the clustering step of the algorithm which determines connected regions of parameter space where the fraction of the ZBP occurrences is larger than $80\%$ on each side.
    Cluster labels are numbers, except ``bg'' is for background, i.e.~no ZBP, and ``no'' is for non-clustered noise.
    (d): Scattering invariant determined numerically in the simulations.
    The comparison with the clustering result illustrates that the data analysis successfully identifies the true topological region as a region of interest, ROI$_1$.
    The red box in (c) and (d) indicates the ROI$_1$ at the lowest $V_p$ --- including a margin of 20\% --- as an example for the input for the next stage.
	\label{fig:phase_1_data_analysis}}
\end{figure}

These steps are illustrated in Fig.~\ref{fig:phase_1_data_analysis}.
Panels (a) and (b) show the probability to find a ZBP on the two sides of the device as a function of plunger voltage and magnetic field.
Fig.~\ref{fig:phase_1_data_analysis}(c) shows the clustered regions of ZBPs, ROI$_1$.
Fig.~\ref{fig:phase_1_data_analysis}(d) shows the scattering invariant computed for the system, with negative values being topological and positive values non-topological.
This last panel is shown only to evaluate the outcome of the analysis on the test data, as the topological invariant is not available directly in experiment.
We note that already the requirement of correlated ZBPs on both ends of the device significantly reduces the parameter space of interest compared to the presence of ZBPs on just one side.
Moreover, it is encouraging to see that there is significant overlap with the most prominent cluster and the presence of the topological phase.

For each ROI$_1$, ranges in field and plunger values that enclose the region are specified as input for the second phase of the TGP.
Due to the slowness of the measurement in the second stage of the TGP, we choose not to scan the cutter gate voltage there.
The cutter gate voltages for stage two are then chosen such that the overlap of the $G_{LL}$ and $G_{RR}$ phase diagram with the average one is maximal.
We note that from analysing simulated data, the ROI$_1$ at the lowest plunger gate voltage tends to be most likely to be topological.
If there is limited time for stage 2, it therefore makes sense to focus on that region first.

\section{Second stage of the TGP}
\label{sec:phase_two}

The input for the second stage of the TGP is the output of the first stage: list of ROI$_1$ -- regions specified by minimal and maximal magnetic field and plunger gate voltage values.
For each ROI$_1$ a cutter value for which the region is most visible is used.
ROI$_1$ is expanded as compared to the most promising cluster by 20\% in all directions to make sure that possible gap closing and ZBP formation features are included in ROI$_1$ [see red box in Fig.~\ref{fig:phase_1_data_analysis}(c,d)].
The full conductance matrix is then measured in these regions and analyzed for the presence of ZBPs and the size of the gap, which is used to define a list of regions of interest, ROI$_2$.

\subsection{Full conductance matrix measurement}

To obtain information about the bulk of the system, we perform a DC measurement of the conductance matrix of the full three-terminal device as a function of the right and left biases.
All the local and non-local ($G_{RR}, G_{LL}, G_{LR}, G_{RL}$) conductances are thereby measured.
Magnetic field, plunger gate, and bias scan steps can be kept the same from the first stage, or the resolution can be increased if the volume of the promising regions is small.
Thus, the data generated in the second stage of the TGP includes a dataset per ROI$_1$.
Each dataset consists of two three-dimensional field-plunger-bias scans, where the bias is scanned separately on the left and the right sides.
In our test example, we have simply re-used the same dataset to execute the stage two analysis of the ROI$_1$ selected by the stage one analysis.
In an experimental run, the measured conductance matrix must be converted into the conductance matrix at the sample before the data analysis, in order to correct for voltage divider effects arising from finite line resistances~\cite{voltagedivider}.

\subsection{Analysis of conductance matrix data}

The input for the data analysis are the conductance matrix traces described above and the outputs are the regions of interest based on both local and non-local conductance -- the ROI$_2$s -- together with the score ascribed to each of them.
The score is a measure of how likely we think the region is truly topological and how large the median gap inside the region is.
We suggest to use a score that is a tuple of the median gap inside the region and the fraction of the boundary of the region that is gapless.
If the percentage of the gapless boundary is high enough -- a threshold of $50\%$ is used in our simulations -- and the median gap inside the region is at least twice as high as the typical electron temperature, we suggest that the region is likely to correspond to an unbroken topological phase and thus the outcome of the TGP is positive, i.e.~we have found a region where MZMs are likely present on the two sides of the device and are protected by the bulk gap.

\begin{figure}[t]
	\includegraphics[width=\columnwidth]{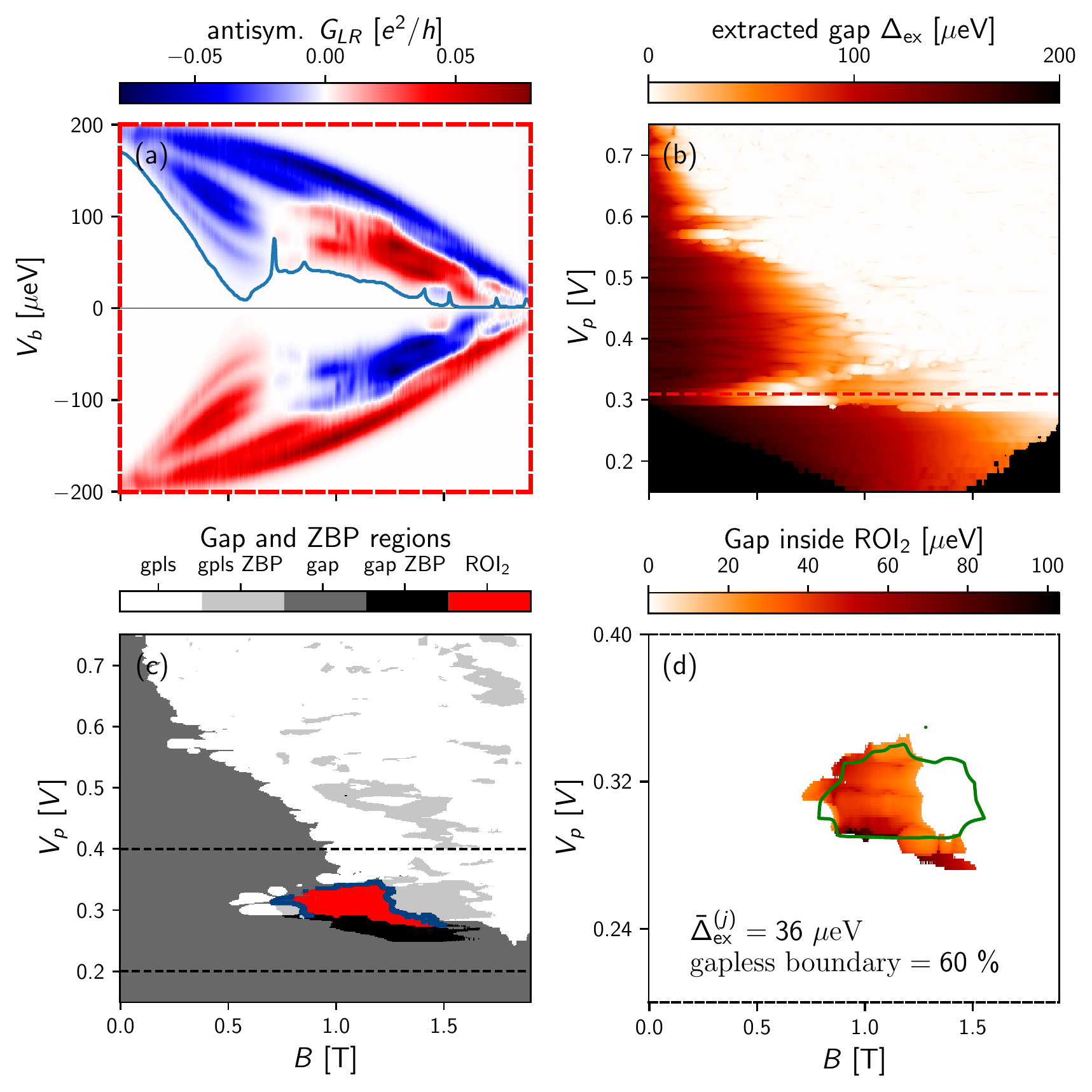}
	\caption{\textbf{Stage two data analysis.}
	The data analysis is illustrated using the same simulation as shown in Fig.~\ref{fig:phase_1_data_analysis}.
	(a) Example of gap extraction from the antisymmetric part of the non-local conductance for a linecut of the phase diagram in (b).
	The determined gap is indicated by the solid blue line.
	For details of the gap extraction algorithm, see App.~\ref{sec:gap_extraction}.
	(b) Gap extracted from the procedure in (a) over the whole phase diagram.
	(c) Phase diagram separated into gapped (gap) / gapless (gpls) regions in presence / absence of ZBPs.
	The identified ROI$_2$ is shown in red and its gapless boundary points are marked blue.
	(d) Zoom-in on the only remaining region of interest ROI$_2$ with the gap plotted only for the points within the region.
	The green line shows the boundaries of the topological region as defined by a scattering invariant $<-0.9$.
	The text labels in (d) indicate the median gap of ROI$_2$ and the fraction of the boundary of the region that is identified as gapless.
	\label{fig:phase_2_data_analysis}}
\end{figure}

To achieve this program, for each ROI$_1$ we do the following:

\begin{enumerate}
	\item Repeat step \ref{item:classify_bias_traces} of the data analysis protocol of Sec.~\ref{sec:data_analysis_one}, in order to verify the measured region is still promising, and potentially to adjust the boundaries of the candidate topological region.
	This is especially important as the output of the ZBP search in stage 1 is averaged over cutter gates while in stage 2 the cutter gate is fixed.
	Additionally, the boundary of the region may have shifted, for example, due to charge redistribution in the device if a long time elapses between the two stages.
	\item Extract the size of the gap $\Delta_{\rm ex}$ for all the points measured in stage 2 by thresholding the non-local conductance.
	The detailed procedure is described in App.~\ref{sec:gap_extraction}.
	\item Determine the parent gap $\Delta_0$ as a function of $B$ at a sufficiently negative voltage where the non-local conductance shows a featureless gap closing.
	\item Define regions of interest ROI$_2$ by clustering points that have ZBPs on both sides of the device (as in stage 1) but additionally show a gap $\Delta_{\rm min}<\Delta_{\rm ex}<\Delta_{\rm max}$.
	Here, $\Delta_{\rm min}$ is our bar for calling a point gapless which we set to \SI{20}{\micro\eV} motivated by the thermal broadening at experimentally achievable electron temperatures of \SI{50}{\milli K}.
	We choose $\Delta_{\rm max}=0.8\Delta_0$ since the topological gap is expected to be a sizable fraction smaller than the parent gap~\cite{stanescu2011majorana}.
	\item Determine the fraction of the boundary of ROI$_2$  which is gapless. A boundary point is called gapless when it is close to a point with $\Delta_{\rm ex}<\Delta_{\rm min}$. In our test data we use a range of 10 pixels, corresponding $\SI{15}{\milli V}$ or $\SI{47.5}{\milli T}$, to define this vicinity.  
	\item Assign a score to the region.
	The score $S$ for region $j$ is a tuple of the percentage of the gapless boundary $p$ and the median extracted gap in the region $\bar{\Delta}_\textrm{ex}$: $S_j = (p_j, \bar{\Delta}_{\rm ex}^{(j)})$.
\end{enumerate}

These data analysis steps are illustrated in Fig.~\ref{fig:phase_2_data_analysis} using the simulated data.
In particular, in Fig.~\ref{fig:phase_2_data_analysis}(a) we show the gap extraction procedure, while  Fig.~\ref{fig:phase_2_data_analysis}(b) depicts the result of the gap extraction for the whole phase diagram.
In Fig.~\ref{fig:phase_2_data_analysis}(c) we show the phase diagram separated into regions that are gapped/gapless and with/without ZBPs according to our analysis.
Moreover, we mark the only ROI$_2$ --- a cluster of points showing ZBPs on both sides and a gap that falls within the window $\Delta_{\rm min}<\Delta_{\rm ex}<\Delta_{\rm max}$ --- by the red color.
The parent gap $\Delta_0$ is extracted at $V_p=\SI{0.24}{V}$, see App.~\ref{sec:trivial_realistic}.
The upper bar for the gap of the ROI$_2$ region is introduced in order to filter out possible regions with trivial ZBPs adjacent to the topological region, see App.~\ref{sec:trivial_realistic}.
We note that the parameter range of the ROI$_2$ region is significantly reduced compared to the regions that are identified by ZBPs alone.
This reduction is especially striking when compared to the almost ubiquitous appearance of ZBPs on only one side of the device, see Fig.~\ref{fig:phase_1_data_analysis}(a,b), and indicates the power of the analysis steps of the TGP.

Fig.~\ref{fig:phase_2_data_analysis}(d) shows a zoom-in on the only remaining region of interest, overlaid with the boundary of the topological region determined from the scattering invariant.
We see that out of the large phase space, the TGP correctly yields a region with large overlap to the only truly topological region in the phase diagram.
Considering the score of the ROI$_2$, $S=(60\%,\SI{36}{\micro eV})$, we also see that in this example the region has a sufficient gap and is predominantly surrounded by a gapless boundary.
The boundary score, however, indicates that not all of the ROI$_2$ is detected to be surrounded by a gap closing.
Fig.~\ref{fig:phase_2_data_analysis}(c) suggests that this is due to a region with trival ZBPs (marked in black) adjacent to the topological region.
Closer examination of the trivial region, see App.~ \ref{sec:trivial_realistic}, reveals that the persistent trivial ZBPs are due to quasi-Majorana states which also obscure the visibility of the phase transition, as discussed in Sec.~\ref{sec:false_negatives}, and thus lead to an underestimation of the fraction of the boundary that is gapless.
It is important to note that such an effect would not produce a false positive of an entirely non-topological phase, but can rather lead to a somewhat misshapen guess for the ROI$_2$ (when trivial regions are included) and possibly false negatives if the quasi-Majorana regions enclose most of the topological region.

Given this observation, further tests and tweaks to the outcome of the TGP can be made to double check and refine the boundaries of a ROI$_2$, especially if the percentage of the boundary which is gapless is close to 50\%.
These tests may for instance include: repeating the scan at higher resolution in plunger gate, magnetic field, and/or bias voltage; repeat the measurement of ROI$_2$ for different cutter gate voltages; scan cutter gate voltages finely for some fixed points in ROI$_2$.

The output of the entire protocol is an estimate for the value of the topological gap in each promising region and its position in the explored parameter space which can be used as a starting point for further experiments with the likely identified MZMs.
In case of multiple ROI$_2$s, different regions could be ranked by a combined score like the product of the median gap inside the region and the fraction of its border that is gapless.

Finally, we note that the above data analysis protocol reflects our current approach.
It is possible there are ways to improve the data analysis of both stages of the TGP and thus we expect the implementation of each step in the data analysis to be subject to continued adjustments.
For more details of the current stage of the data analysis and the choice of the approaches used, see App.~\ref{app:data_analysis}. We also note that while our current set of analysis parameters is intentionally chosen to be restrictive for identifying an ROI$_2$ as topological, a more lenient version of the TGP might be implemented by relaxing some of the analysis parameters.

\section{Conclusions}
\label{sec:conclusions}
In this paper we put forth a protocol -- the topological gap protocol (TGP) -- that consists of a set of measurements and data analysis steps with the goal to quickly and reliably screen a device for the presence of a topological superconducting phase. The enhanced reliability of the TGP stems from holistically considering the characteristics of the topological \emph{phase} and the accompanying phase transition, rather than merely identifying points or small pockets in parameter space that are consistent with a Majorana interpretation.

The measurement consists of a combination of local and non-local conductance measurements taken over a large range of the plunger gate voltage and magnetic field values.
In the analysis we identify regions of interest (ROI$_1$) in parameter space where ZBPs are present at both ends of the wire.
We then further investigate these regions by extracting the transport gap from the non-local conductance.
Gap measurements identify the topological phase transition, rather than the topological phase itself, and provide a powerful tool in combination with the local conductance information.
The output of the TGP is another set of regions of interest (ROI$_2$), given by the intersection between the ZBP regions and the gapped regions, together with a score quantifying the median gap of the regions and the fraction of their boundary that is gapless.
A ROI$_2$ with a sizable gap and a predominantly gapless boundary is very likely to be topological.

The TGP has significant advantages over previous methods of identifying MZMs: (1) The well-defined procedure removes the subjectivity that is inherent in the notion of a stable ZBP.
(2) The requirements for a positive outcome of the TGP are much more stringent than the simple presence of ZBPs and thus significantly reduces the chance of a false positive identification of an MZM.
(3) Since the TGP involves scans over a large parameter space it acts as a tool to both search for and verify MZM candidates.
As such it can be more efficient in delivering a verdict  whether a given device is promising from the perspective of MZMs than ad-hoc searches for ZBPs.

The drawbacks of the TGP are that the measurements are more involved and that, despite being superior to purely local analysis tools, the outcome of the TGP is not perfect.
Smooth potentials at both ends of the wire can lead to non-topological ZBPs, thereby giving birth to regions of interest ROI$_1$ that extend outside the topological phase.
At the same time, they can obscure the gap closing, thereby making it difficult to find the topological phase transition.
More generally, a spatially slowly-varying potential can cause the system to undergo the phase transition at different magnetic fields in different parts of the system (e.g.~middle and near the ends).
These types of effects can give rise to false negatives --- wrongly labeling a topological region surrounded by trivial ZBPs as entirely non-topological.
Our simulation results, however, indicate that it is unlikely that the entire topological phase is enclosed by trivial ZBPs.
Moreover, through device design, modelling and improvements in growth and fabrication it should be possible to minimize such device inhomogeneities.
A different type of false negative for identifying the presence of a topological phase can occur if the potential near the ends of the wire pushes the MZMs too far from the leads to be visible as ZBPs.
From the perspective of identifying accessible MZMs this case, however, is actually a true negative even though a topological phase may exist inside the wire.

Overall, for our parameter choice, the TGP is intentionally biased to give increased confidence in the positive outcomes, while accepting that some false negatives might be produced.
A region that is positively evaluated by the protocol thus likely indicates the presence of an accessible topological phase. As such, the TGP provides an excellent starting point for more elaborate MZM experiments and topological quantum computation.

\section*{Author contributions}
BvH, CN, DIP, EAM, GdL, JDW, LC, MaT, RML, SH, TK conceived the TGP.
BvH, BN, DIP, EAM, TK, created the working version of the TGP using numerical simulations performed by GWW and TL, and data analysis help from MiT and VS.
EAM, GdL, JDW, LC, and SH performed experiments clarifying the approach.
The figures were prepared by BN, DIP, EAM, TK, and TL.
The manuscript was written by BvH, CN, DIP, EAM, RML, TK with input from all authors.

\acknowledgements
We would like to thank 
Andrey~Antipov, Bela~Bauer, Tom~Dvir,  Eoin~O'Farrell, John~Gamble, Esben~Bork~Hansen, Victor~Hartong, Andrew~Higginbotham, Jonne~Koski, Leo~Kouwenhoven, Charlie~Marcus, Signe~Markussen, and Karl~Petersson   for many conversations in relation with this project, Sergei~Gronin, Geoff~Gardner, Ray~Kallaher, and Michael~Manfra for growing the material used to demonstrate RF-DC conversion, and Maren~Kloster and Shivendra~Upadhyay for fabricating the devices.

\nocite{apsrev41Control}
\bibliographystyle{apsrev4-1}
\bibliography{TGP_refs}
\newpage
\appendix

\section{Details of simplified model simulation}
\label{sec:details_toy_model_simulation}

We use a simplified one-dimensional Rashba wire model to generate test data for the TGP.
The system is defined in Eqs.~(\ref{eq:H_toy1}-\ref{eq:H_toy2}), where
the superconductor is integrated out into a self-energy boundary condition.
This process is well-justified in practice because the superconductor is typically very disordered and thus boundary effects and geometrical resonances are completely smeared out.
One can thus use the bulk Green's function for the disordered s-wave superconductor, which is reproduced by coupling the semiconductor to superconducting 1D chains~\cite{rosdahl2018andreev}.
Numerically, we thus attached to every lattice site of the discretized Hamiltonian that is connected to the superconductor a semi-infinite one-dimensional superconductor.
The effective semiconductor-superconductor coupling parameter $\gamma_0$~\cite{stanescu2011majorana} is chosen such that the induced gap is slightly smaller than the parent superconductor gap --  92\% in our case.
In the self-energy we use a small broadening parameter $\eta$ by substituting $\omega \rightarrow \omega + i \eta$.
We discretize the Hamiltonian and calculate the quantum transport numerically using the package Kwant~\cite{groth2014}.
The specific parameters used are listed in Tab.~\ref{tab:simulation_parameters}.

\section{Details of realistic geometry simulation}
\label{sec:details_realistic_simulation}

\begin{figure}
    \centering
    \includegraphics[trim={0 0 80 0},clip,width=0.98\columnwidth]{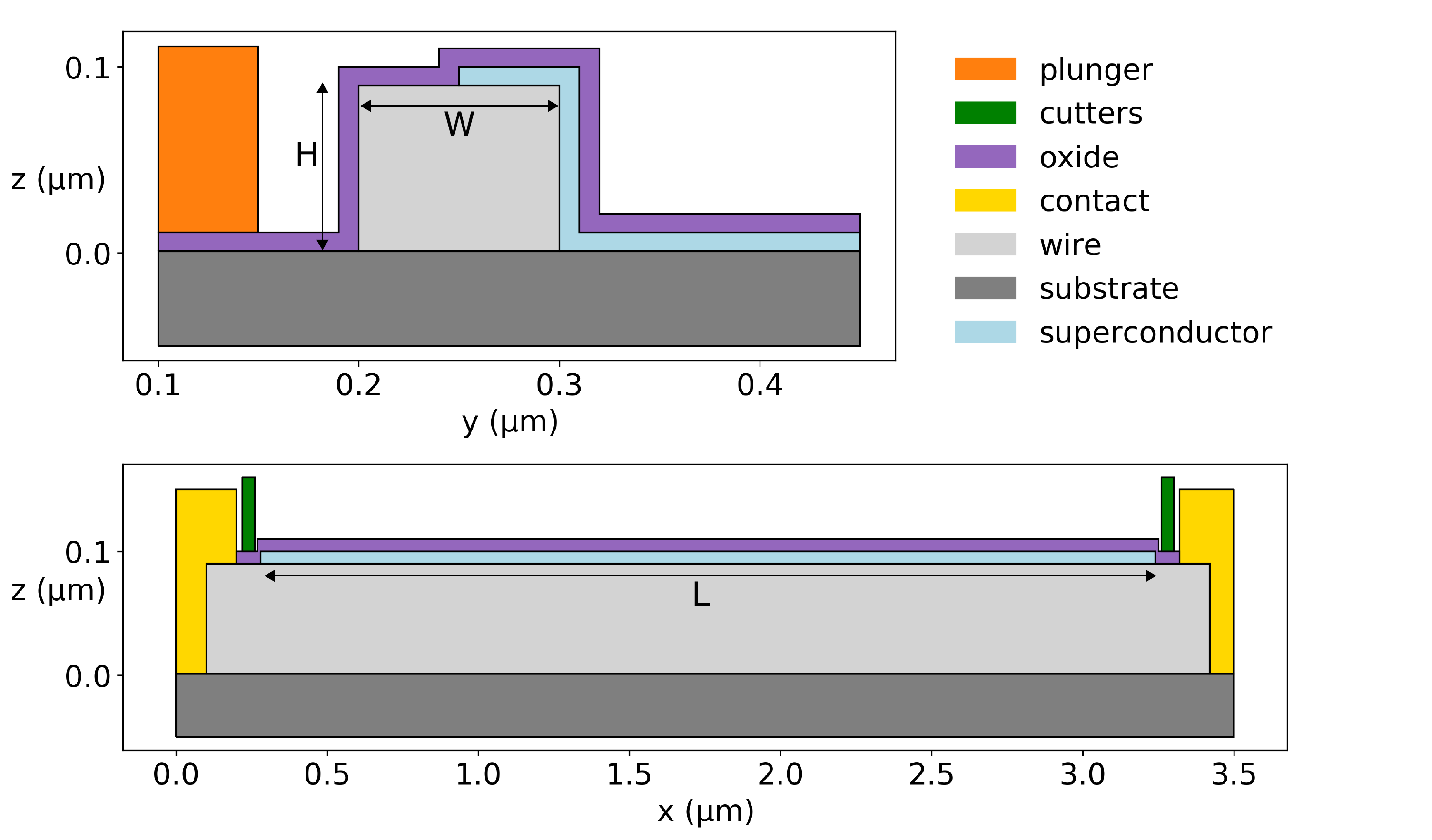}
    \caption{\textbf{Device modelled in the realistic geometry.}
    Cross-sectional and longitudinal cut of the model used in the realistic geometry simulations: substrate (GaAs, dark grey), wire (InSb, light grey), ohmic contacts (Au, gold), cutter gates (Au, orange), plunger gate (Au, green), aluminum superconductor (Al, blue) and oxide ($\mathrm{HFO_2}$, pink).
    The parameters used to simulate the materials in the electrostatic calculation and tight-binding simulation can be found in Tab.~\ref{tab:simulation_parameters}.
    The dimensions of the proximitized region are $L=\SI{3}{\micro\meter}$, $W=100$ nm, $H=90$ nm.
    The thickness of the oxide and superconductor layer are both 10 nm.
    The lattice spacing used in the tight-binding simulation is 10 nm.}
    \label{fig:realistic_geometry}
\end{figure}

\begin{figure*}
    \centering
    \includegraphics[width=0.85\textwidth]{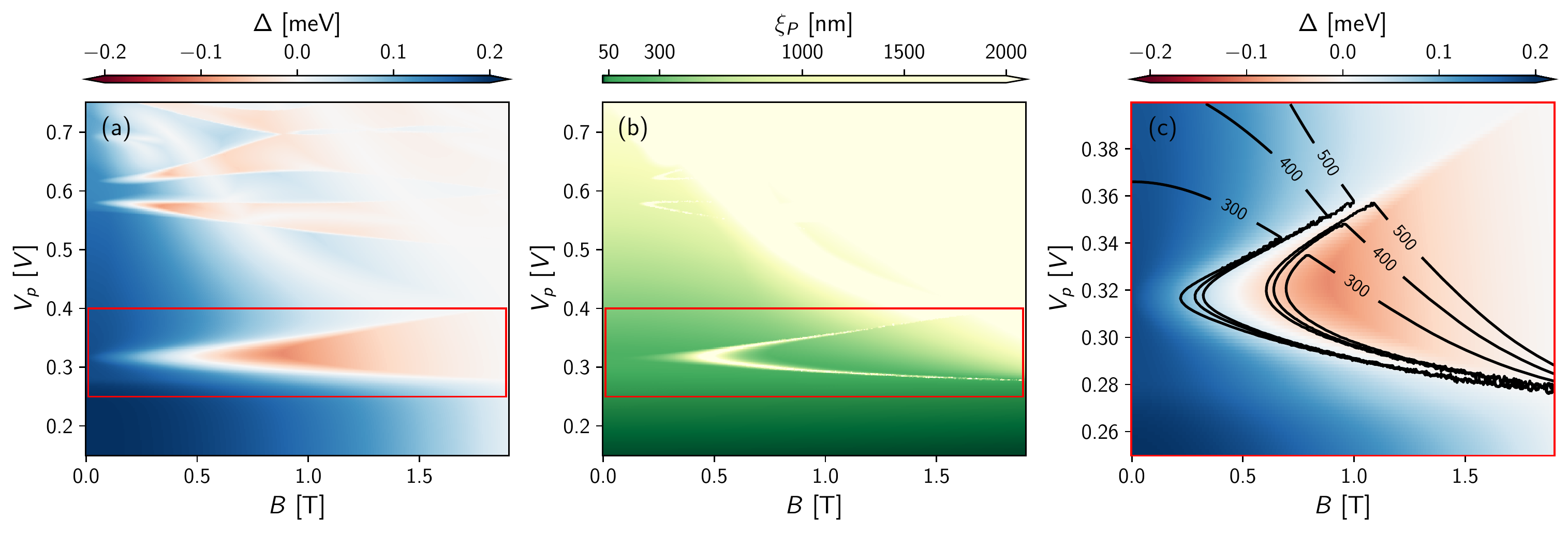}
    \caption{\textbf{Phase diagram and coherence length for the three-dimensional simulation of an InSb/Al wire.}
    All quantities were extracted from the band structure computed from the 2D cross-section of the system in the middle of the device, see Fig.~\ref{fig:realistic_geometry}.
    The gap in (a) and (c) is multiplied by the sign of the Pfaffian such that a negative gap energy corresponds to a topological gap.
    The superconducting coherence length $\xi_P$ in (b) is computed by computing the wavelength of the slowest decaying mode.
    The color scale is cut off at $\SI{2000}{\nano\meter}$ to emphasize the change of the coherence length within the topological phase.
    The red box in (a) and (b) indicates the parameter range of (c).
    The contour lines in (c) indicate fixed levels of $\xi_P$ in nm.
    }
    \label{fig:coherence_length}
\end{figure*}

Beyond the simulation of one-dimensional simplified models, we have performed advanced three-dimensional simulations of a proximitized nanowire in order to generate a mock dataset of a protocol instance and produce the data analysis illustrated in Fig.~\ref{fig:phase_1_data_analysis} and Fig.~\ref{fig:phase_2_data_analysis} of the main text.
For this purpose we have chosen to simulate an InSb nanowire with a rectangular cross-section, proximitized by an Al layer on its top facet and on one of its side facets.
The detailed geometry of the device used in the simulation is shown in Fig.~\ref{fig:realistic_geometry}.
The simulations solve for the electrostatic potential in the Thomas-Fermi approximation.
The superconductor is included at the level of a local, energy-dependent self-energy, inducing $s$-wave pairing at the interface sites, analog to the simplified model simulations.
The simulation code has been already described in detail in Refs.~\cite{vaitiekenas2020fullshell, shen2020parity, Kringhoj2021}.
The conductance matrix was again computed numerically using the package Kwant~\cite{groth2014}.
Our numerical calculation includes the phenomenological addition of a small broadening parameter $\eta$ to the self-energy to guarantee convergence.
In addition we convolve the resulting conductance traces with the derivative of a Fermi function to model temperature broadening.
The simulation parameters listed in Tab.~\ref{tab:simulation_parameters} are consistent with previous simulations of InSb/Al nanowires~\cite{shen2020parity}.
Fig.~\ref{fig:coherence_length}(a) shows the topological phase diagram of the wire in the clean limit.

\begin{table}[]
	\centering
	\begin{tabular}{|l|l|l|}
		\hline
		Parameter                                 & Simplified model      & 3D sim.               \\
		\hline
		Effective mass of semicond.               & 0.015 $m_e$              & 0.0135 $m_e$       \\
		Rashba spin-orbit strength                & 100 meV nm            & 150   meV nm          \\
		semiconductor $g$-factor                  & -30                   & -50                   \\
		Band offset (w.r.t InSb)                  & ---                   & 50 meV                \\
		Interface coupling $\gamma_0$               & 2.87 meV              & ---                   \\
		Aluminum effective mass                   & $m_e$                 & $m_e$                 \\
		Aluminum gap $\Delta_0(B=0)$              & 0.25 meV              & 0.2   meV             \\
		Aluminum critical field                   & $\infty$              & 2   T                 \\
		Aluminum Fermi energy                     & 10 eV                 & 10 eV                 \\
		Broadening $\eta$ in $\Sigma_\mathrm{SC}$ & $\SI{0.1}{\micro\eV}$ & $\SI{0.1}{\micro\eV}$ \\
		Temperature broadening                    & $\SI{4.5}{\micro\eV}$ & $\SI{4.5}{\micro\eV}$ \\
		\hline
		Material name/ Parameter                  &                       &                       \\
		\hline
		GaAs                                      &                       &                       \\
		\; relative permittivity                  & -                     & 13.1                  \\
		InSb                                      &                       &                       \\
		\; relative permittivity                  & -                     & 16.8                  \\
		\; direct bandgap                         & -                     & 235 meV               \\
		\; electron affinity                      & -                     & 4590 meV              \\
		\; electron mass                          & -                     & 0.0135 $m_e$          \\
		\; density interface traps                & -                     & 0                     \\
		HfO2                                      &                       &                       \\
		\; relative permittivity                  & -                     & 25.0                  \\
		Au                                        &                       &                       \\
		\; workfunction                           & -                     & 5285 meV              \\
		vacuum                                    &                       &                       \\
		\; relative permittivity                  & -                     & 1.0                   \\
		        
		\hline
	\end{tabular}
	\caption{Parameters used in the simplified model and realistic geometry three-dimensional simulations of a proximitized nanowire.
	For the electrostatics calculation of the realistic geometry model, the substrate (GaAs) was modeled as a dielectric.
	Here $m_e$ is electron mass.}
	\label{tab:simulation_parameters}
\end{table}

Our simulation also includes disorder at the level of a random, uncorrelated onsite potential sampled from a box distribution.
Disorder is only added to the semiconductor sites between the two cutter gates, and the disordered potential is added only after the Thomas-Fermi calculation.
Our treatment of disorder is simplified, and aims at introducing some non-ideality in the simulation and the resulting phase diagram rather than modeling impurities in a real device.
The simulation reported in the main text has a disorder with strength 10 meV.
The corresponding mean free path can be estimated using the simple formula $U=\sqrt{3\pi / l_e {m^*}^2 a^3}$, which is valid for a bulk three-dimension system, and which yields $l_e\approx 3000$~nm (here, $l_e$ is the mean free path, $m^*$ the effective mass and $a=10$~nm the lattice spacing).
While the mean free path in the simulated multi-band wire will deviate from this value and depend on magnetic field and gate voltage, this order-of-magnitude estimate guarantees that the disorder employed in the simulations is weak enough to not completely disrupt the presence of a topological phase.
This is confirmed by the calculation of the scattering invariant shown in Fig.~\ref{fig:phase_2_data_analysis}(d) and also by the direct computation of the coherence length \cite{Nijholt2016} in the clean limit shown in Fig.~\ref{fig:coherence_length}(b) and (c).
The figure shows that the coherence length can be as short as 300 nm in the topological phase occurring around $V_\textrm{p}=\SI{0.3}{\volt}$, much smaller than our estimate of the mean free path.

We have generated an extensive dataset of conductance matrix values, covering all the relevant parts of the topological phase diagram of the device.
To verify the topological character of the system, we computed the scattering invariant $\mathcal{Q}=\det r$, where $r$ is the reflection matrix from either of the two normal leads~\cite{akhmerov2011, fulga2011scattering}.
Our numerical calculations were performed using the Adaptive~\cite{nijholt2019} package for an efficient sampling of the parameter space and then interpolated on a dense grid for the execution of the data analysis stages of the TGP.

\section{Details of data analysis}
\label{app:data_analysis}

\subsection{ZBP extraction methods}

\begin{figure}[h]
    \centering
    \includegraphics[width=0.95\linewidth]{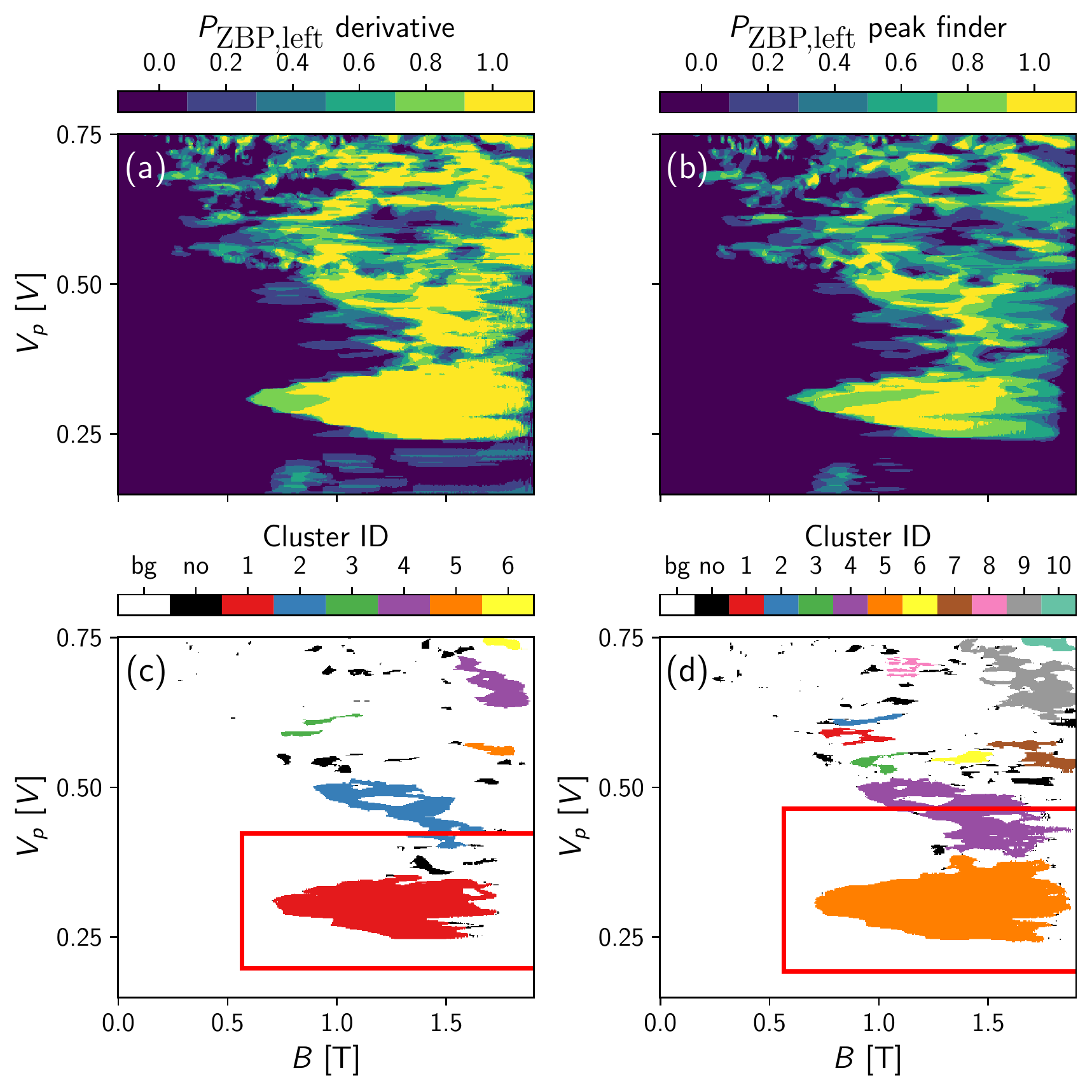}
    \caption{\textbf{Zero bias peak extraction methods.} Comparison between ZBP extraction methods.
    (a) and (c) are identical to Fig.~\ref{fig:phase_1_data_analysis} (a) and (c) respectfully.
    In particular, (a) shows the fraction of the cutter values that show ZBPs on the left side of the device where the ZBPs are extracted using the second derivative method and (c) shows the clustered data.
    (b) shows the fraction of the cutter values that show ZBPs on the left side of the device where the ZBPs are extracted using the peak finding method and (d) shows the clustered data.
    Comparing (a) and (b) or (c) and (d) we see that the ZBP extraction methods are largely equivalent.
    The red box in (c) and (d) indicates the cluster with the lowest $V_p$, including a margin of 20\%.}
    \label{fig:app_zbp_extraction_methods}
\end{figure}

In this section we compare two methods of the ZBP extraction that can be used in the TGP: direct peak finding and a criterion based on the second derivative of the conductance versus bias voltage~\cite{fornieri2019evidence}.
We show that for simulated data they produce largely equivalent results.
The input data for each ZBP extraction method is the local conductance measured as a function of the applied bias, $G_{LL}(V_L)$ and $G_{RR}(V_R)$.

The first method uses the standard peak finding routine of the \verb|scipy| Python library~\cite{SciPy-NMeth}.
To weed out spurious peaks and we apply thresholds on the peak height and prominence (relative to all the peaks in the same trace).
Application of this method to experimental data also benefits from applying a Savatzky-Golay filter to the conductance trace, to mitigate the effect of noise.
Peaks that are within temperature broadening from zero bias, i.e.~$3.5k_B T\approx \SI{15}{\micro\eV}$ are considered zero bias.

The second method exploits the second derivative of the conductance near zero bias, using the idea that a ZBP will be associated with a negative curvature of the conductance trace at zero bias.
More concretely, we determined the presence of a ZBP if the second derivative of the conductance at zero bias is negative with absolute value larger than $\frac{e^2}{h} \left(\SI{50}{\micro\volt}\right)^{-2}$, corresponding to the curvature of a ZBP with 0.5 $\frac{e^2}{h}$ falling off to zero over at a bias of $\SI{50}{\micro V}$.
The derivative is computed over a bias window of $\SI{30}{\micro\volt}$.
We used the method of the second derivative for the ZBP detection in the main text.
For comparison, we replot the ZBPs from the the main text Fig.~\ref{fig:phase_1_data_analysis} in in Fig.~\ref{fig:app_zbp_extraction_methods}(a) side by side with the other peak finding method shown in Fig.~\ref{fig:app_zbp_extraction_methods}(b).
Panels (c) and (d) compare the corresponding results of the clustering of regions with ZBPs on both sides of the device.

Visually the two methods agree very well.
The correlation coefficient between the two datasets in Fig.~\ref{fig:app_zbp_extraction_methods}(a,b) is 89.9\%.
Additionally, we checked that the ROI$_{1,2}$ do not change significantly when using the different methods.
We therefore conclude that any of them can be used, however, the second derivative method is faster in practice because it only requires processing the data in a small window around zero bias.

\subsection{Clustering of ROI$_1$}

\begin{figure}[h]
    \centering
    \includegraphics[width=0.95\linewidth]{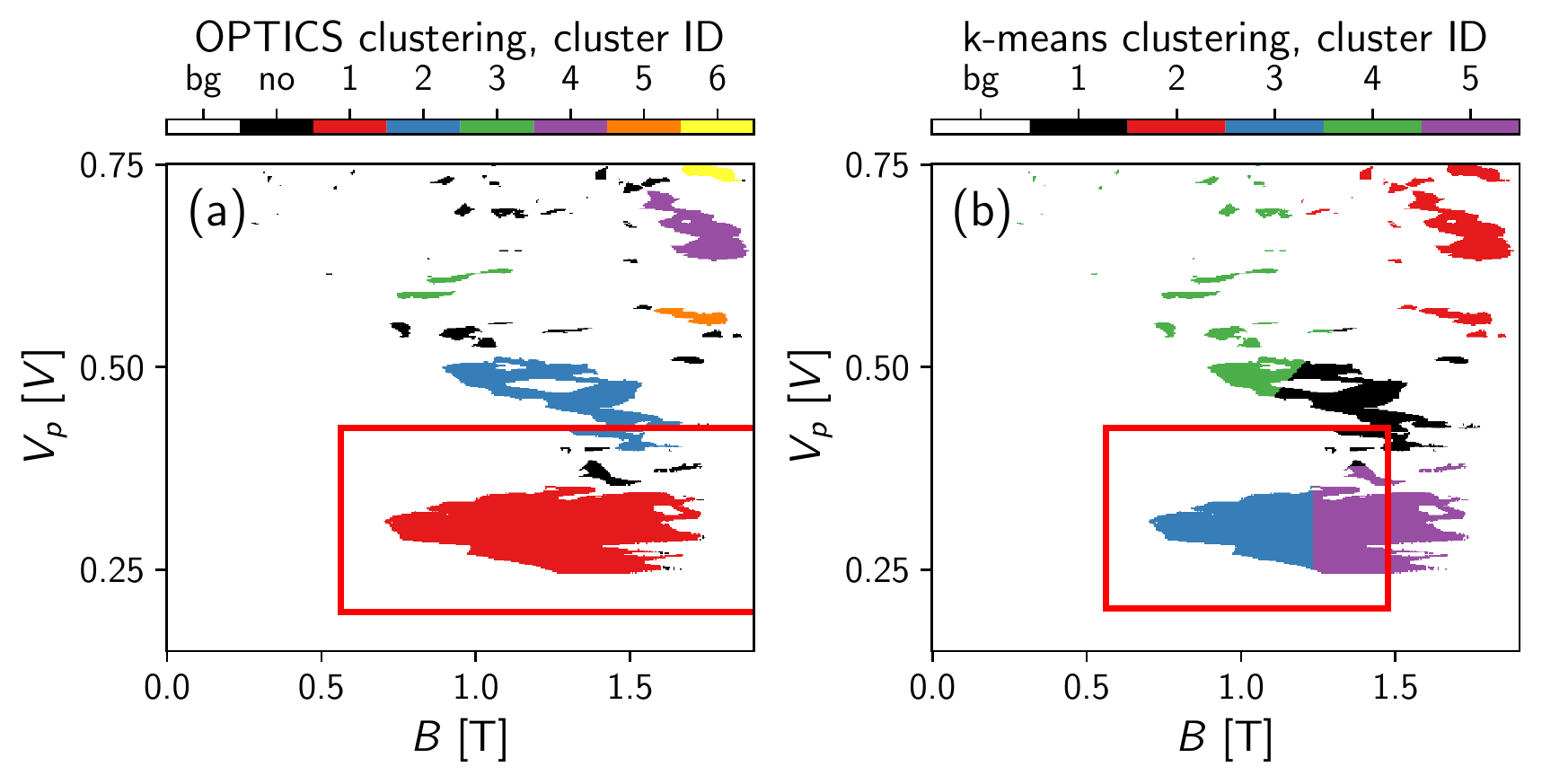}
    \caption{\textbf{Clustering methods.}
    (a) shows the default clustering method we used \texttt{OPTICS}.
    The continuous regions are correctly identified and too small clusters are neglected as noise.
    (b) uses the conventional clustering algorithm \texttt{KMeans} with 5 clusters.
    We see that the methods fails to determine the optimal separation of the data into continuous regions.
    The red box in both subplots indicates the cluster with the lowest $V_p$, including a margin of 20\%.}
    \label{fig:app_clustering_methods}
\end{figure}

To obtain the ROI$_1$ in the first stage of the TGP we use the density-based clustering algorithm \texttt{OPTICS} from \texttt{scikit-learn} python library~\cite{scikit-learn}.
The input for the algorithm is the collection of points
where the system shows a ZBP on both sides.
The output is a set of all the clusters of these points.
The clusters are formed by \texttt{OPTICS} recursively: a new point can be added to the cluster if the distance from the new point to some of the points in the cluster is smaller than a threshold value times the average distance between the points within the cluster.
Finally, by specifying how far in absolute units the points in a cluster can be, \texttt{OPTICS} can classify random one-off points as noise -- in our case they can correspond to trivial ABSs crossing zero energy where the lack of stability indicates the absence of a topological origin.
We used the \texttt{OPTICS} routine with parameters \texttt{xi=0.1}, \texttt{min\_cluster\_size=350}, \texttt{min\_samples=10}, and \texttt{max\_eps=10}.
With the resolution of our data the minimal cluster size corresponds to an area of $\SI{50}{\milli T}\times\SI{50}{\milli V} $ in the magnetic-field/plunger-voltage space.

Importantly, the shape of the clusters obtained by \texttt{OPTICS} is not constrained -- it can be concave or convex without affecting the clustering at all.
This is because the clusters are determined by the distances within each cluster and between any cluster and nearby points.
This was an important consideration in the choice of the clustering method.
For example, the standard clustering algorithm, \texttt{KMeans}, can produce only linear boundaries of the clusters and generically does not produce good ROI$_1$ boundaries.
The comparison between the two clustering methods is shown in Fig.~\ref{fig:app_clustering_methods}.

\subsection{Gap extraction algorithm}
\label{sec:gap_extraction}
The gap extraction algorithm is based on thresholding the value of the non-local conductance.
Simple thresholding based on 1D bias traces at fixed magnetic field and plunger voltage can be difficult due to the varying strength of the non-local signal.
For example, there can be bias traces with overall very small non-local conductance which makes relative thresholding based entirely on such a trace unreliable.
This problem becomes particularly pronounced when working with experimental data, where it can be beneficial to consider a bias window smaller than the zero field induced gap to save time for the data acquisition.

To resolve some of these issues we extract the gap over the parameter space by considering 2D bias-field cuts, see Fig.~\ref{fig:phase_2_data_analysis}(a).
Since the gap collapses at high fields even considering a small bias window will result in an appreciable non-local conductance in at least some part of the 2D cut.
We then take the maximal absolute value of the non-local conductance throughout the 2D cut as a reference point to set the threshold of all the 1D bias scans within the 2D cut at a fraction of that reference point.
In Fig.~\ref{fig:phase_2_data_analysis}(a) we use a fraction of 1\% for this thresholding.

To further optimize the gap extraction procedure, especially anticipating experimental noise, it can be useful to apply certain filters to the non-local conductance data of the 2D cuts before performing the thresholding.
We found it useful to apply a small-range median filter (to catch outliers), antisymmetrize the non-local conductance (to fix the value at zero bias to zero) and apply a small-range Gaussian filter at the end (to avoid sharp jumps in the field-dependence of the gap).
The algorithm used for the data analysis of the simulated data also includes these filters but they are set to a range of just a few pixel which doesn't have much of an effect on the finely sampled data.

\section{Trivial ZBPs in realistic geometry simulation data.}
\label{sec:trivial_realistic}

\begin{figure}[]
    \centering
    \includegraphics[width=0.95\columnwidth]{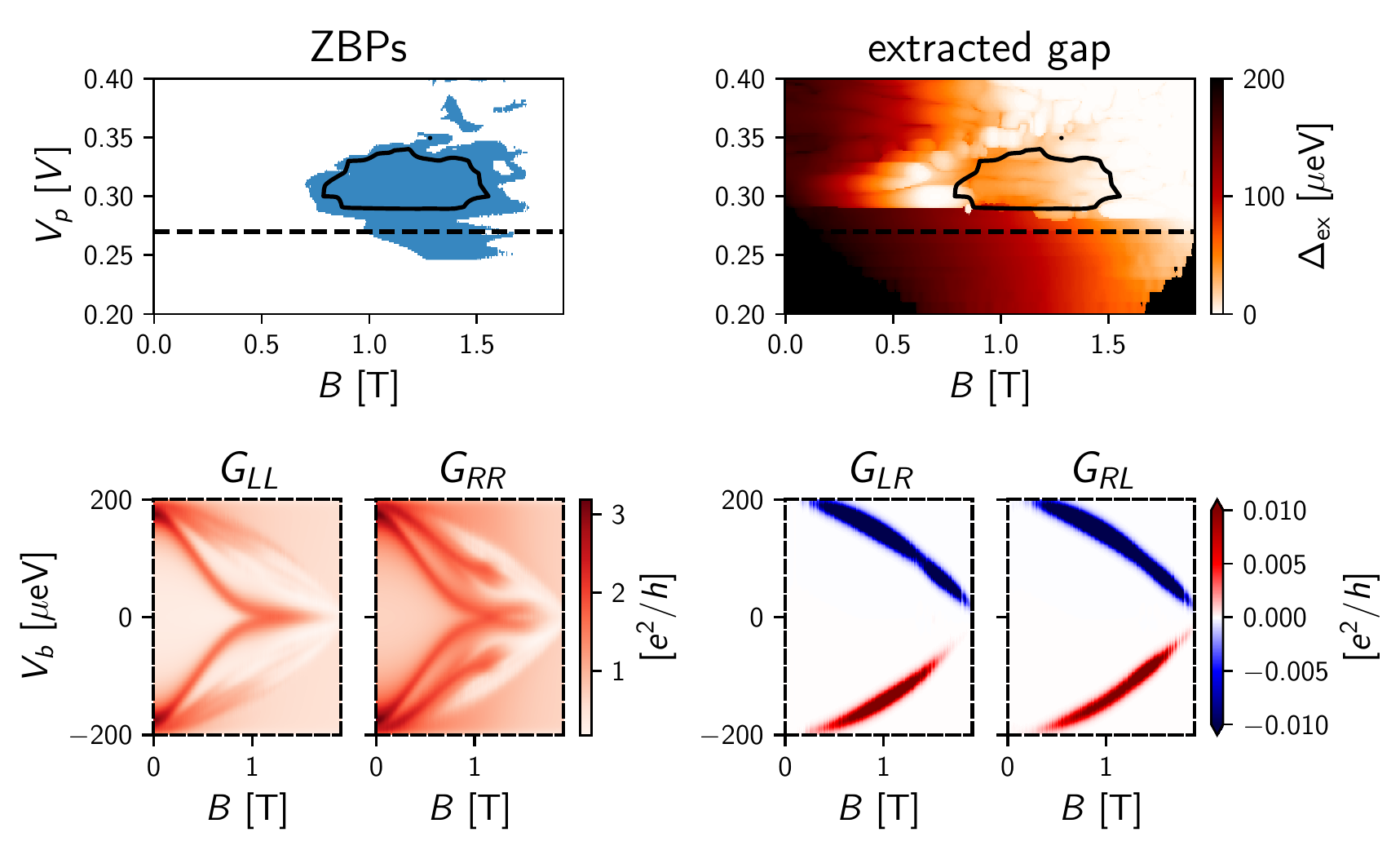}
    \caption{\textbf{Trivial ZBPs in realistic geometry model}.
    Same data as in Fig.~\ref{fig:phase_2_data_analysis} of the main text.
    The first row shows the presence of ZBPs on both sides of the device and the extracted gap.
    The black contour line marks the topological phase boundary as extracted by the scattering invariant.
    The black dashed line indicates a cut at $V_p=\SI{0.27}{\milli V}$ with the corresponding conductance matrix plotted in the second row.
    The local conductance shows the presence of trivial ZBPs while the non-local condutance is featureless except for the closing of the parent gap.}
    \label{fig:trivial_realistic}
\end{figure}

Figure~\ref{fig:phase_2_data_analysis}(c) shows an extended region with trivial ZBPs on both sides of the device (maked in black) that is adjacent to the topological region.
To investigate these ZBPs further we provide a line cut with the full conductance matrix at $V_p=\SI{0.27}{V}$ in Fig.~\ref{fig:trivial_realistic}.
We observe the presence of ZBPs without any features in the non-local conductance, except for the closing of the parent gap $\Delta_0$ which thus indicates that the ZBPs are of trivial origin.
The presence of trivial ZBPs near the topological phase is reminiscent of the quasi-Majorana modes discussed in Sec.~\ref{sec:false_negatives}.
Since the simulated device is symmetric and these peaks are correlated we thus attribute them to smooth potentials at the end of the device rather than disorder-induced peaks.

We also checked that there is a very faint gap-closing feature ($\sim 10^{-6}e^2/h$) at $V_p=\SI{0.28}{\milli V}$, which is close to the lower boundary of the topological phase identified by the TGP and the scattering invariant.
Already at the value $V_p=\SI{0.27}{\milli V}$ plotted in Fig.~\ref{fig:trivial_realistic} this gap closing feature is gone (even at high resolutions $\sim 10^{-6}e^2/h$) which indicates that the gap quickly reopened into a trivial phase.
Similar to the discussion of the quasi-Majorana case in Sec.~\ref{sec:false_negatives}, the smooth potential itself suppresses the visibility  and the phase transition in this example would likely not be detectable at the experimentally available resolution.

Fortunately, the featureless non-local conductance allows for a way to exclude such trivial ZBPs from the identified ROI$_2$.
It is clear from Fig.~\ref{fig:trivial_realistic} that the extracted gap cannot be the topological gap which has to be a sizable fraction smaller than the parent gap, see e.g.~\cite{stanescu2011majorana}, due to the low $g$-factor and the absence of spin-orbit coupling in the parent superconductor aluminum.
The absence of a detectable topological gap can be caused by two effects: (1) The considered point in parameter space is simply not topological or (2) the visibility to the modes of the topological band is severely suppressed by the shape of the potential at the end of the system.
As mentioned in Sec.~\ref{sec:false_negatives} these two effects actually often occur together in the case when quasi-Majorana modes are present in the vicinity of the topological phase.
In either case both effects do not constitute an accessible topological region the TGP is designed to identify and we thus exclude these regions by requiring the gap in ROI$_2$ to be smaller than a certain $\Delta_{\rm max}$ which is a large fraction of the parent gap.
Here we use $\Delta_{\rm max}=0.8\Delta_0$ and extract the parent gap from the cut at the lowest $V_p$ at which a featureless gap closing can be extracted without visibility issues.
The latter can appear in the simulations far below the chemical potential of the lowest band as indicated by the black regions in the plot of the extracted gap.
We chose $V_p=\SI{0.24}{\milli V}$ for the extraction of the parent gap as it gives acceptable visibility while lying outside the region of detected (trivial) ZBPs.
The corresponding plots of the non-local conductance look similar to the one shown in Fig.~\ref{fig:trivial_realistic}.

\section{Further examples of Rashba wire simulations}
\label{sec:examples_rashba_wire}

\begin{figure*}[h!]
    \centering
    \includegraphics[width=0.92\textwidth]{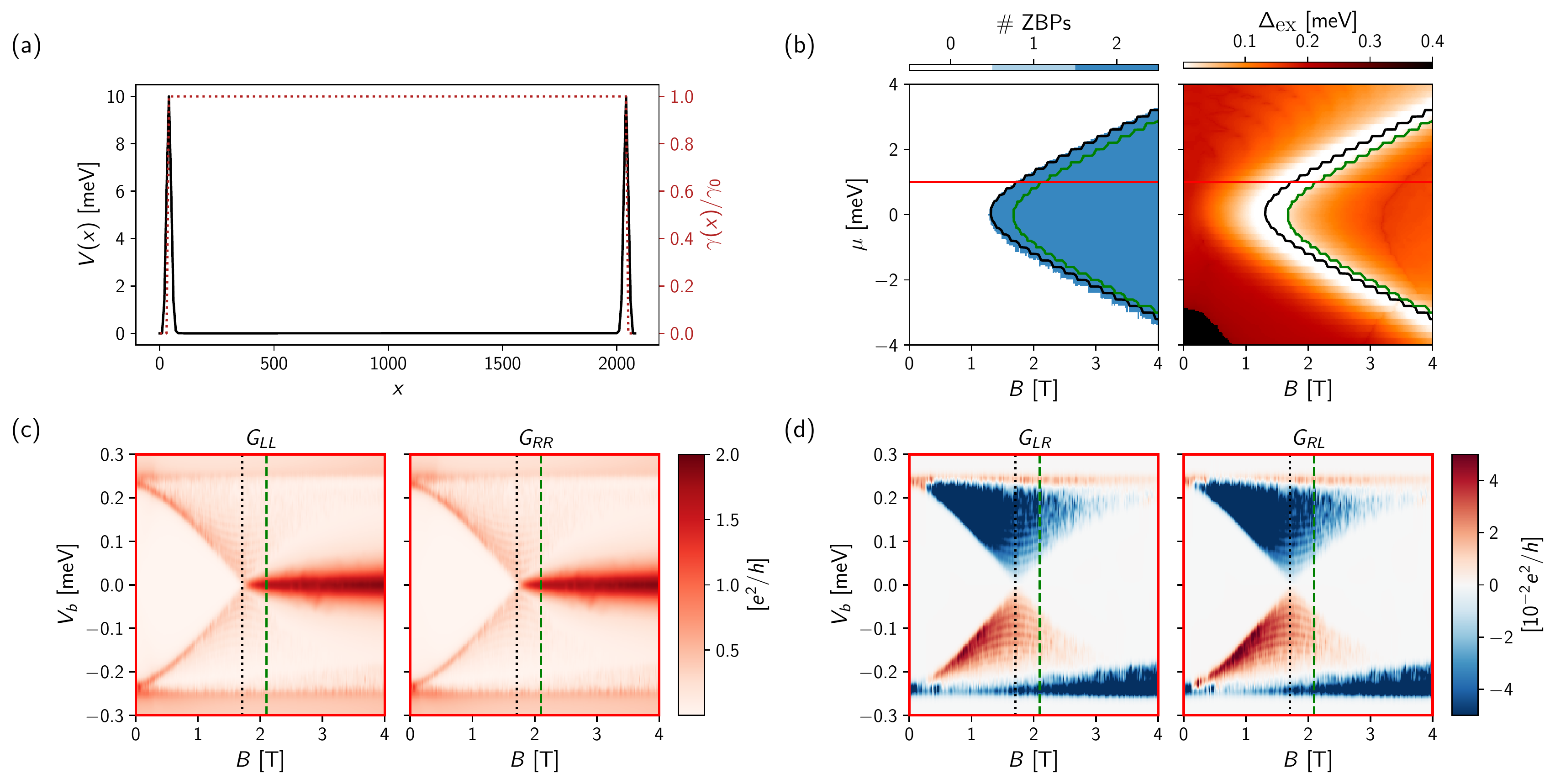}
    \caption{\textbf{Simplified model: Clean case.}
    Superconductor-semiconductor coupling and potential shown in (a).
    Phase diagrams obtained from ZBP extraction and gap extraction from the conductance matrix in (b).
    The solid black line indicates a topological phase as identified by a scattering invariant $<-0.9$, while the green line indicates a region with ZBPs on both ends and a gap $>\SI{20}{\micro\eV}$ (corresponding to ROI$_2$, see Sec.~\ref{sec:topogap_brief_description}).
    The median gap of the green region is $\SI{114}{\micro\eV}$ with $100\%$ of its boundary determined to be gapless.
    Local (c) and non-local (d) conductance examples at the chemical potential marked in (b).
    The vertical black dotted and green dashed lines in (c) and (d) indicate the intersection with the lines of corresponding color in (b).}
    \label{fig:app_clean}
\end{figure*}

\begin{figure*}
    \centering
    \includegraphics[width=0.92\textwidth]{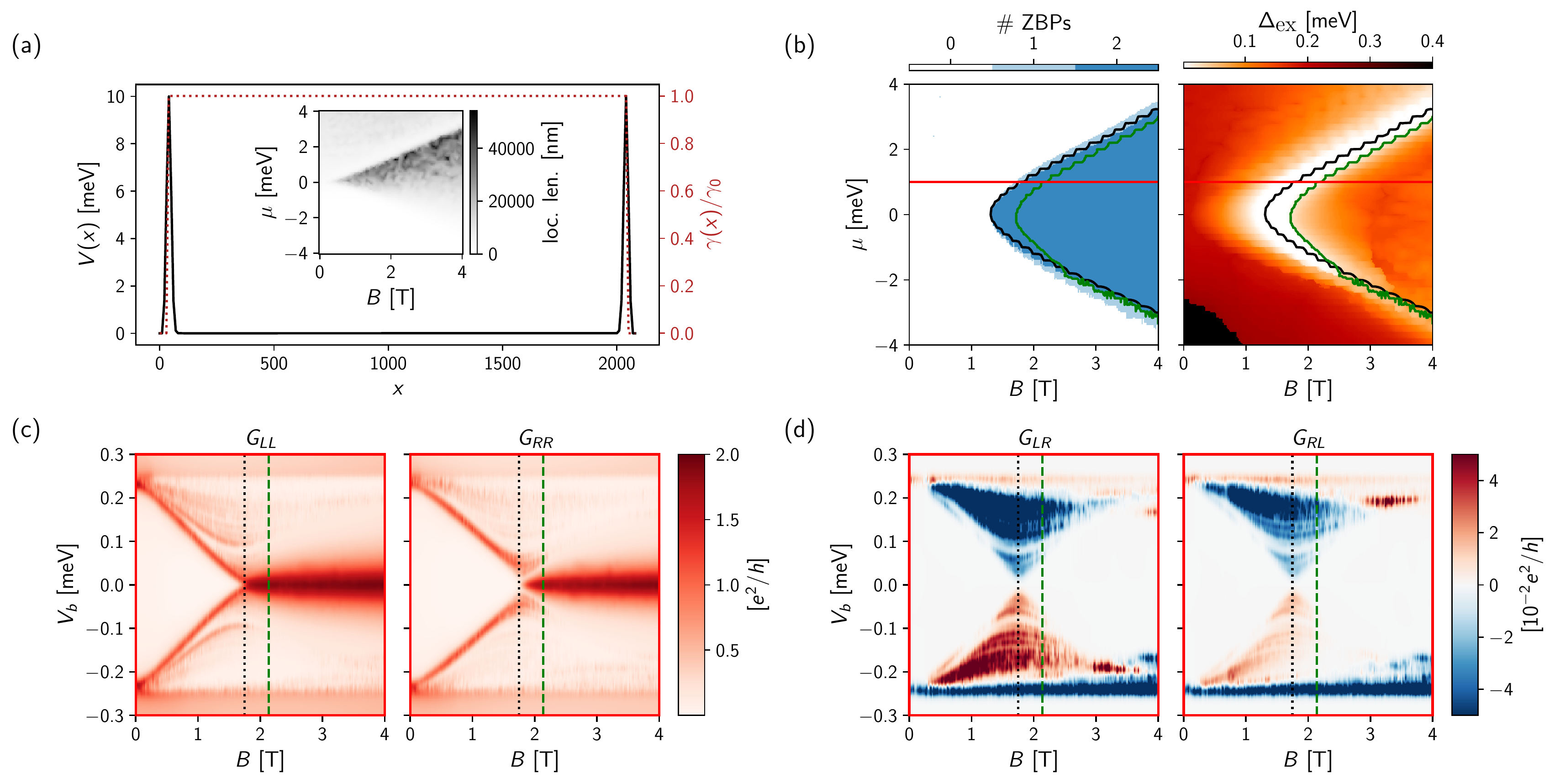}
    \caption{\textbf{Simplified model: Weak disorder.}
    Superconductor-semiconductor coupling and smooth part of the potential shown in (a).
    The inset shows the normal state localization length of the semiconductor as a function of $B$ and $\mu$.
    For comparison the minimal topological coherence length in the clean system is $128$ nm.
    Phase diagrams obtained from ZBP extraction and gap extraction from the conductance matrix in (b).
    The solid black line indicates a topological phase as identified by a scattering invariant $<-0.9$, while the green line indicates a region with ZBPs on both ends and a gap $>\SI{20}{\micro\eV}$ (corresponding to ROI$_2$, see Sec.~\ref{sec:topogap_brief_description}).
    The median gap of the green region is $\SI{116}{\micro\eV}$ with $68\%$ of its boundary determined to be gapless.
    Local (c) and non-local (d) conductance examples  at the chemical potential marked in (b).
    The vertical black dotted and green dashed lines in (c) and (d) indicate the intersection with the lines of corresponding color in (b).}
    \label{fig:app_weakdisorder}
\end{figure*}

\begin{figure*}
    \centering
    \includegraphics[width=0.95\textwidth]{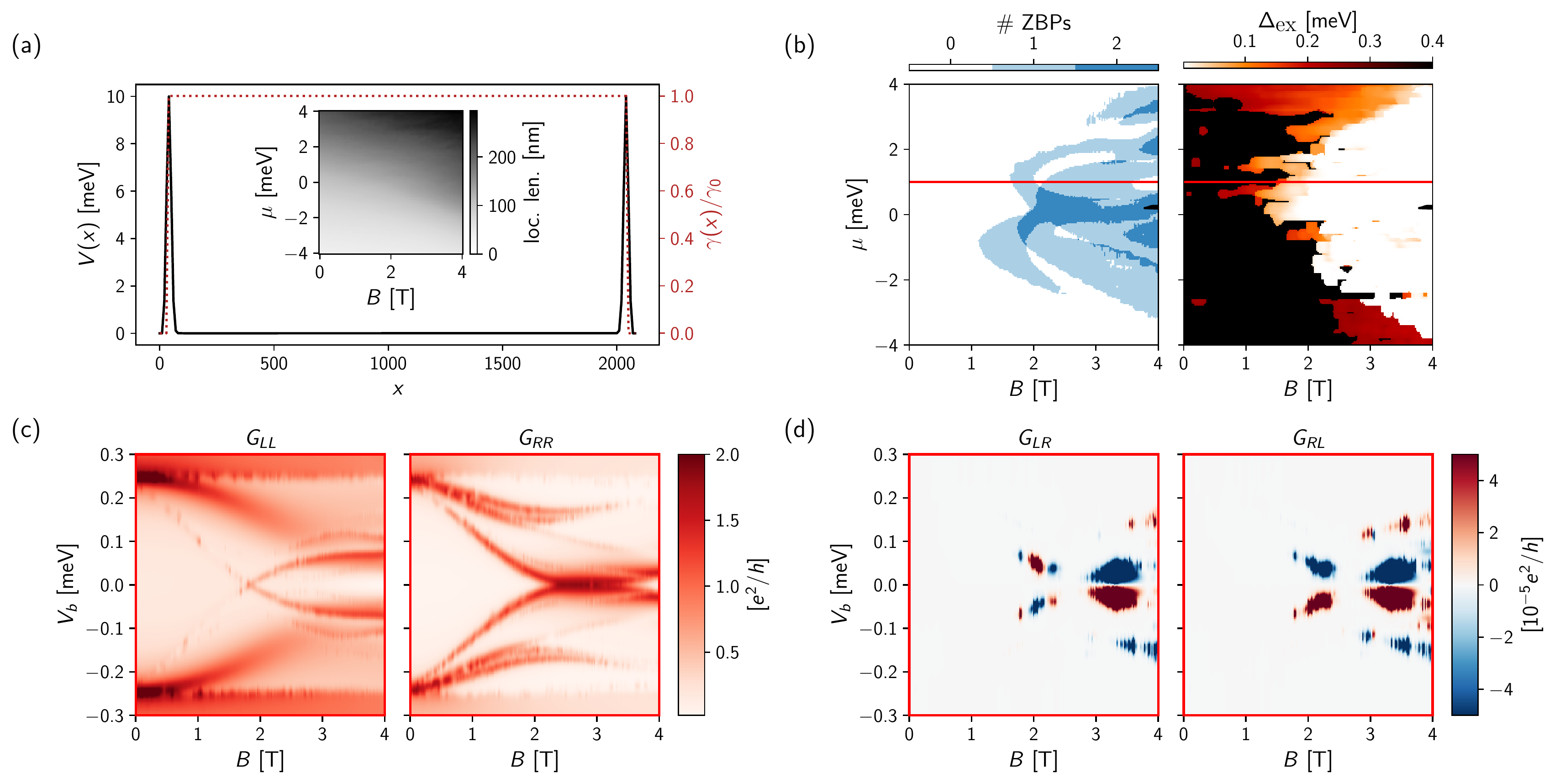}
    \caption{\textbf{Simplified model: Strong disorder.}
    Same plots as in Fig.~\ref{fig:app_weakdisorder} but with stronger disorder as indicated by the inset of (a) which shows that the typical localization length does not exceed the minimal clean coherence length of \SI{128}{\nano m}.
    No sizable ROI$_2$ is detected.}
    \label{fig:app_strongdisorder}
\end{figure*}

\begin{figure*}
    \centering
    \includegraphics[width=0.95\textwidth]{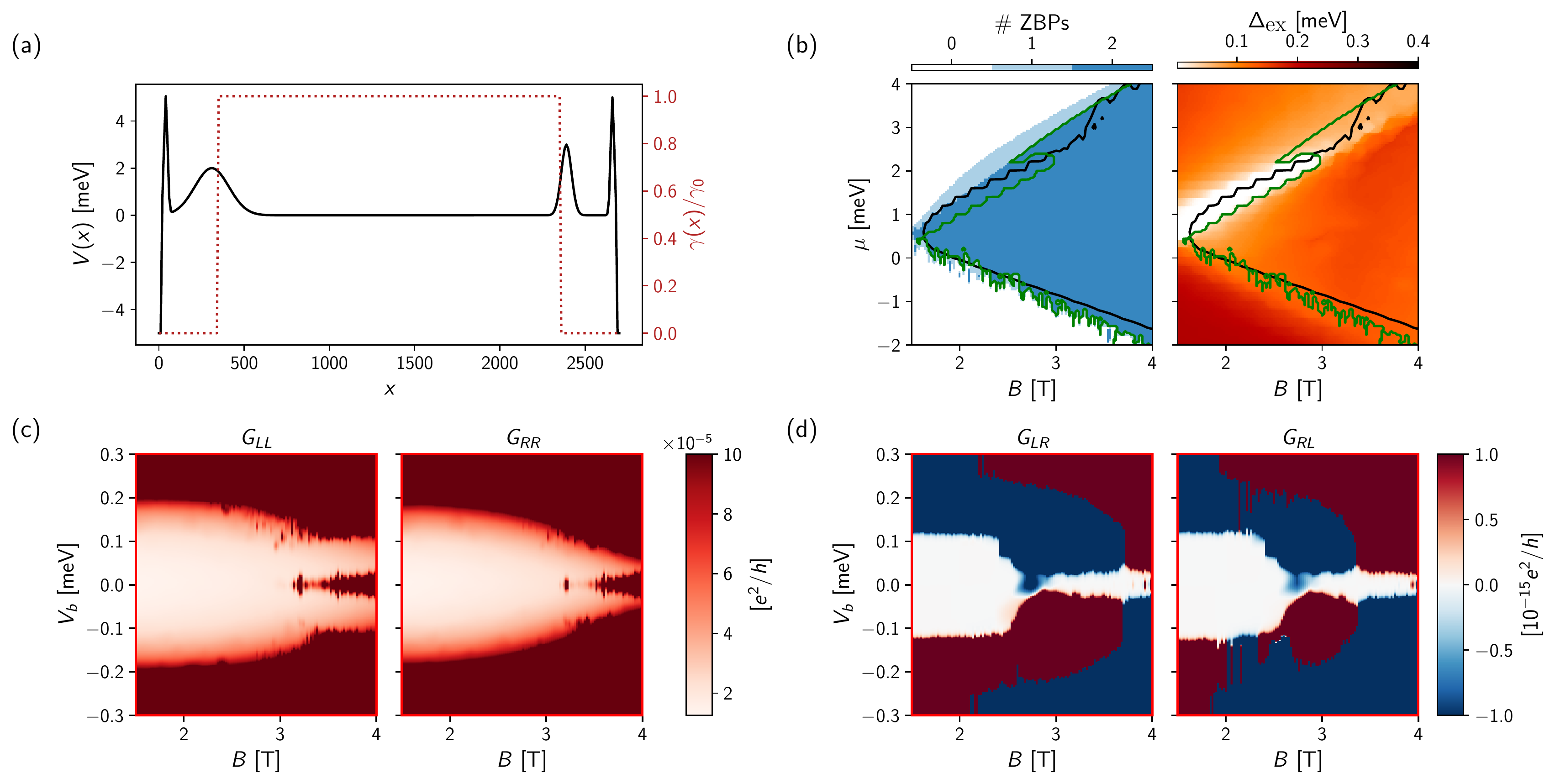}
    \caption{\textbf{Simplified model: Quasi-Majorana scenario at negative chemical potential.}
    Same plots as in Fig.~\ref{fig:app_2QM} but with (c-d) at a $\mu=-2$ meV and with a vastly over-saturated colorscale. This is an illustration of the gap closing which is invisible in any realistic observation.}
    \label{fig:app_2QM-2}
\end{figure*}

\begin{figure*}
    \centering
    \includegraphics[width=0.95\textwidth]{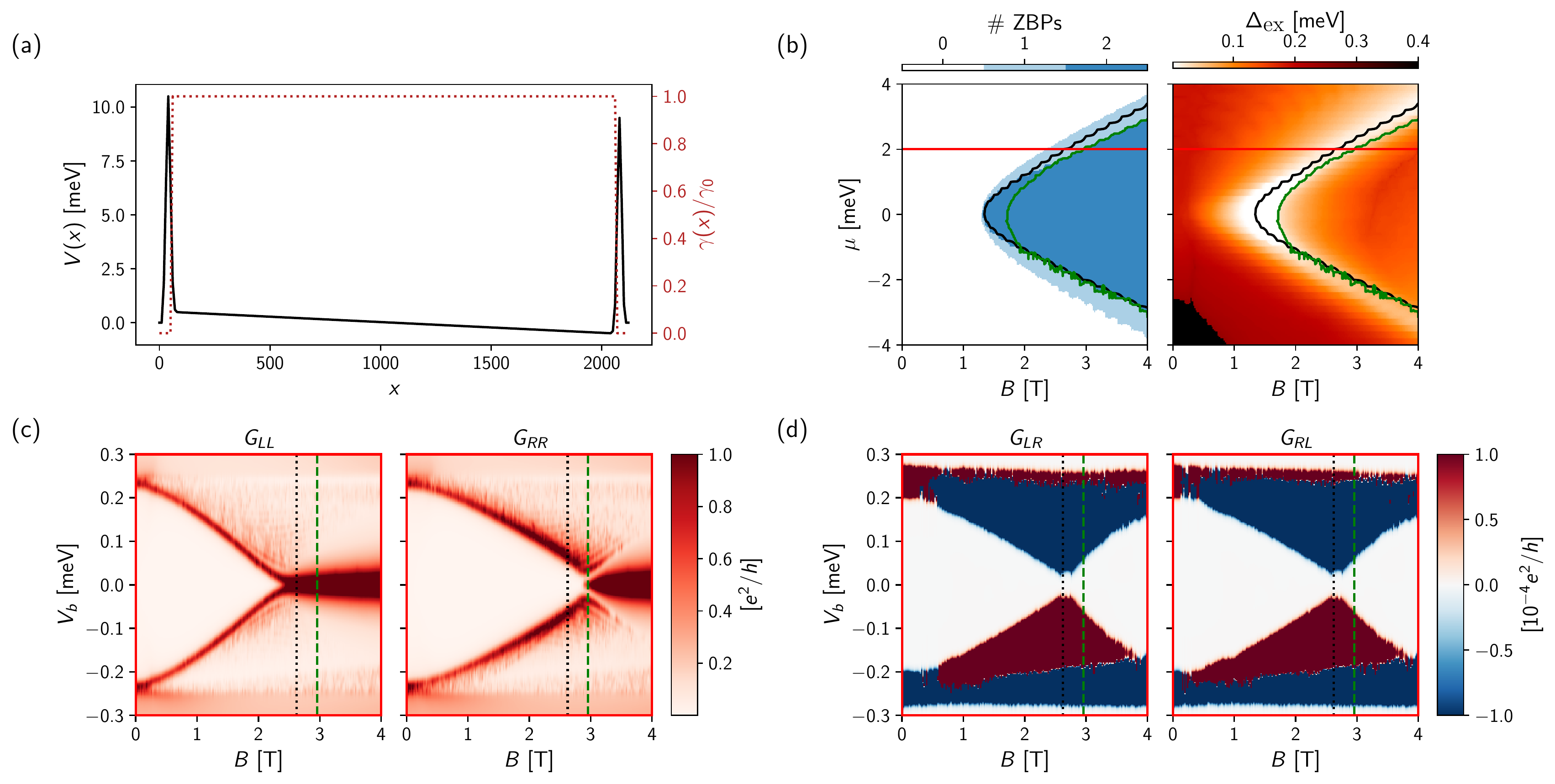}
    \caption{\textbf{Simplified model: Potential gradient.} Same plots as in Fig.~\ref{fig:app_clean} but in the presence of a potential gradient shown in (a).
    Phase diagrams obtained from ZBP extraction and gap extraction of conductance matrix in (b).
    Local (c) and non-local (d) conductance examples  at the chemical potential marked in (b).
    The median gap of the green region is $\SI{118}{\micro\eV}$ with $55\%$ of its boundary determined to be gapless.
    The vertical black dotted and green dashed lines in (c) and (d) indicate the intersection with the lines of corresponding color in (b).}
    \label{fig:app_gradient}
\end{figure*}

\begin{figure*}
    \centering
    \includegraphics[width=0.95\textwidth]{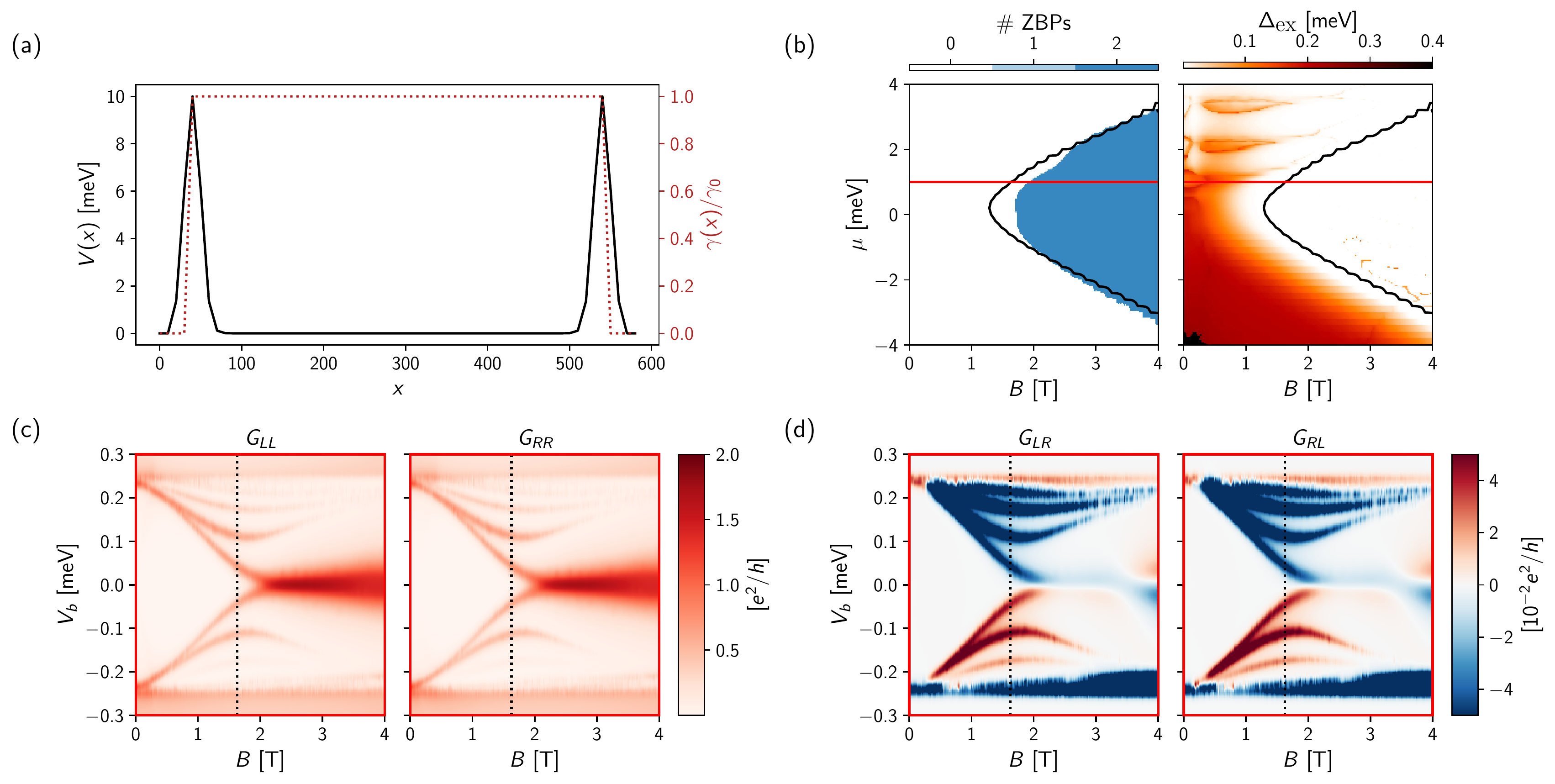}
    \caption{\textbf{Simplified model: Short system.} Same plots as in Fig.~\ref{fig:app_clean} but for a short system of length $\sim 500$ nm, see (a).
    For comparison the minimal topological coherence length of the system is 128 nm.
    Phase diagrams obtained from ZBP extraction and gap extraction of conductance matrix in (b).
    Local (c) and non-local (d) conductance examples  at the chemical potential marked in (b).
    No sizable ROI$_2$ is detected.
    The vertical black dotted line in (c) and (d) indicates the intersection with the black line in (b).}
    \label{fig:app_short}
\end{figure*}

Here we list some more examples of the simulations of a one-dimensional Rashba wire discussed in Sec.~\ref{sec:false_negatives}.
Figures~\ref{fig:app_clean} to \ref{fig:app_short} are all presented in the same format to facilitate the comparison of different scenarios: namely, we show the potential profile used in the simulations [panel (a)], 2D maps of zero bias peak and extracted gap [panel(b)], and the simulated field vs bias scans of local [panel(c)] and non-local [panel(d)] components of the conductance matrix at a representative value of the chemical potential.

To begin with, Fig.~\ref{fig:app_clean} shows the ideal case of a clean wire: as expected the ROI$_2$ closely follows the scattering invariant phase diagram.
Figures \ref{fig:app_weakdisorder} and \ref{fig:app_strongdisorder} show how the clean case is modified in the presence of weak and strong disorder.
In the strong disorder case, there is no longer a sizable region of interest at the second stage of the TGP in agreement with the absence of topological regions as identified by the scattering invariant.

Figure~\ref{fig:app_2QM-2} shows an over-saturated cut at $\mu=-2$ meV in the same regime of smooth end of wire potentials discussed in Sec.~\ref{subsec:smooth_side}.
The extreme over-saturation is only for theoretical illustrative purposes and allows to see otherwise invisible features of the topological phase.
Specifically, there is a ZBP appearing around $B=3$ T in the local conductance and a gap closing and reopening feature around the same value of the magnetic field in the non-local conductance.

Figure~\ref{fig:app_gradient} shows the performance of the TGP data in the presence of a potential gradient. The local conductance clearly shows the onset of ZBPs at different magnetic fields due to the potential gradient.
The strength of the gradient is on the border of what the TGP is still registering as a positive outcome.
Depending on the precision of the non-local conductance measurement such a case might already be labeled as a false positive in practice.

Finally, Fig.~\ref{fig:app_short} shows the case of a short system.
In the case of significant below-gap transmission through the system due to the system length not significantly exceeding the coherence length, the non-local conductance can no longer be used to reliably extract the gap of the system.
In the example here the finite-size signal in the local-conductance makes the region of topological scattering invariant look gapless which leads to a vanishing size of the region of interest in stage 2 of the TGP.
Since such strongly overlapping Majorana modes lack many of the features of MZMs this case should not be considered as a false negative but as a correct identification of the absence of well separated MZMs.

\end{document}